\documentstyle[amsmath,amssymb,graphicx,rotate]{article}
\def\a{\hbox{{\fontfamily{ptm}\selectfont\fontshape{it}
\fontseries{b}\selectfont \hspace{-.5mm}[21] }}}

\def\b{\hbox{{\fontfamily{ptm}\selectfont\fontshape{it}
\fontseries{b}\selectfont \hspace{-.5mm}[3]\,}}}
\def\c{\hbox{{\fontfamily{ptm}\selectfont\fontshape{it}
\fontseries{b}\selectfont \hspace{-.5mm}[111]\,}}}
\def\d{\hbox{{\fontfamily{ptm}\selectfont\fontshape{it}
\fontseries{b}\selectfont \hspace{-.5mm}[2]\,}}}
\def\e{\hbox{{\fontfamily{ptm}\selectfont\fontshape{it}
\fontseries{b}\selectfont \hspace{-.5mm}[11]\,}}}
\def\f{\hbox{{\fontfamily{ptm}\selectfont\fontshape{it}
\fontseries{b}\selectfont \hspace{-.5mm}[1]\,}}}

\def\be{\begin{eqnarray}}
\def\ee{\end{eqnarray}}
\def\nn{\nonumber}

\def\l[{\phantom.\!\![}
\def\R{{\mathfrak R}} 

\def\fR{{\cal R}}
\def\HH21{^*{\cal H}^{4_1}_{[21]}(A|\,q)}
\def\SS21{^*\!S_{[21]}}
\def\H{{\cal H}}
\def\HH{{\cal H}^{4_1}}
\def\SS{S^*}

\def\AL{{\cal A}}

\hoffset= - 1.0in         

\title{{\bf Knot polynomials in
the first non-symmetric representation}
\vspace{.5cm}}
\author{{\bf A.Anokhina}\footnote{ {\small {\it
MIPT, Dolgoprudny, Russia} and {\it ITEP, Moscow, Russia}};
anokhina@itep.ru}, \ {\bf A.Mironov}\footnote{ {\small {\it Lebedev
Physics Institute}; {\it ITEP, Moscow, Russia} and {\it IIP,
Federal University of Rio Grande do Norte, Natal, Brazil}}; mironov@itep.ru;
mironov@lpi.ru}, \ {\bf A.Morozov}\thanks{{\small {\it ITEP, Moscow,
Russia} and {\it IIP,
Federal University of Rio Grande do Norte, Natal, Brazil }}; morozov@itep.ru}, \ {\bf And.Morozov}\thanks{{\small {\it
Moscow State University} and {\it ITEP, Moscow, Russia}};
Andrey.Morozov@itep.ru}\date{ }}

\begin{document}

\setcounter{footnote}{3}

\setcounter{tocdepth}{3}

\maketitle

\vspace{-6.5cm}

\begin{center}
\hfill FIAN/TD-28/12\\
\hfill ITEP/TH-52/12\\
\hfill IIP-TH-29/12
\end{center}

\vspace{5cm}

\begin{abstract}
We describe the explicit form and the hidden structure of the
answer for the HOMFLY polynomial for the figure eight
and some other 3-strand knots
in representation $\a$. This is the first result for non-torus
knots beyond (anti)symmetric representations, and its evaluation is
far more complicated. We provide a whole variety of different
arguments, allowing one to guess the answer for the figure eight
knot, which can be also partly used in more complicated situations.
Finally we report the result of exact calculation for figure eight and some other 3-strand knots
based on the previously developed sophisticated technique of multi-strand
calculations. We also discuss a formula for the superpolynomial
in representation $\a$ for the figure eight knot, which heavily
relies on the conjectural form of superpolynomial expansion nearby
the special polynomial point. Generalizations and details will be
presented elsewhere.
\end{abstract}

\bigskip

\bigskip

\section{Introduction}

The theory of knot polynomials is today at the crossroads
between numerous well developed subjects:
conformal field theory,
non-perturbative Yang-Mills and Seiberg-Witten theory,
AGT relations,
topological models, integrable systems
and, of course, the Chern-Simons and knot theories {\it per se}.
Most unresolved problems in all these fields get
concentrated around the basic unanswered questions
about knot polynomials and their dependencies on numerous
natural variables.
Further development in these areas depends heavily on
availability of explicit expressions for knot polynomials,
which can help to distinguish between generic and particular
properties that can be accidentally valid for rather
simple knots and representations.
However, explicit calculations in knot theory are well known
to be quite sophisticated, and non-trivial examples are
rather difficult to evaluate.
Still, there is a lot of progress in this direction
during the last years.
This note reports a new progress in the still
intractable direction: evaluation of knot polynomials \cite{ACJK,Wit,HOMFLY} in representations
with the Young diagrams which contain more than one row or column.
The celebrated Rosso-Jones formula \cite{RJ} allows one
to get these formulas for arbitrary torus knots and links, but
nothing is yet known beyond this class.
In this note we report the answer for HOMFLY polynomials of
the figure eight knot
$4_1$ and some other 3-strand knots in the first non-trivial representation $\a$.

This does not necessarily look like a big step:
these are the simplest non-toric knots (and the figure eight knot belongs to the simple class
of twist knots),
representation belongs to the simple class of hook diagrams.
But today it is really on the border of unknown:
different calculational approaches developed so far, reach
the same level of complexity for this example, and this
complexity is at the level of nowadays potential of
publicly available (not specially dedicated) computer
facilities (using the software like MAPLE or Mathematica).
For example, the cabling approach requires dealing with
the $9$-strand braids in the fundamental representation \cite{AMMM},
while the direct approach to the colored HOMFLY polynomials {\it a l\'a} \cite{IMMM3}
requires the knowledge of $9\times 9$ mixing matrices,
which are still beyond the advanced list of \cite{IMMM3}.
Explicit knowledge of this new HOMFLY polynomials sheds
a new light on the general properties
of HOMFLY and superpolynomials, confirms some and
discards other existing hypotheses.

We return to these implications in separate detailed
publications, where extensions of this result in
various directions are also discussed.
Here we just collect the arguments, directly applying
them to the simplest example of the HOMFLY polynomial
${\cal H}^{4_1}_{[21]}(A|q)$. These arguments allow us to guess the answer,
which we confirm later by the direct calculation. In fact, basically the same calculation provides
the HOMFLY polynomials in representation $\a$ not only for the figure eight knot, but
for all 3-strand knots, and we list the first few examples in the Appendix. At last, we
discuss different extensions of the answer for ${\cal H}^{4_1}_{[21]}(A|q)$ to
any hook diagrams and to the superpolynomial.

We do not go into details of any of the arguments
leaving them for dedicated texts. It is just amusing how many stories
are brought together at this particular small crossroad.

Throughout the paper, we use the notation
\be
\{x\}=x-{1\over x},\ \ \ \ \ \epsilon=q-q^{-1},\ \ \ \ \ \ \ [n]_q={q^n-q^{-n}\over q-q^{-1}}
\ee
in the latter case we omit the index $q$ unless it can lead to a misunderstanding.

\section{Generic ingredients of formulas for twist knots
\cite{IMMMfe,twist}
\label{gentwi}}

In \cite{IMMMfe} we described the general structure of
HOMFLY and superpolynomials for $4_1$ knot in all
symmetric and anti-symmetric representations.
Our answers were straightforwardly generalized in \cite{ind,Fuji} to
all twist knots.
In fact, these formulas provide a spectacular expansion
in the DGR-like "differentials" \cite{DGR},
and, as will be explained in \cite{twist},
their true {\it raison d'etre}
is intimately related to the {\it evolution method}
of \cite{DMMSS}, which is directly applicable to the
twist knots.

The main ingredients of the construction of \cite{IMMMfe}
can be summarized as follows:

\begin{itemize}
\item
With each box of the Young diagram one associates the $Z$-factor
\be
Z_{I|J}(A|q) = \{Aq^I\}\{Aq^{-J}\}
\ee
By default $(I,J)$ are not quite the coordinates $(i,j)$ of the box
in the diagram, rather $I=2(l_j-i)+1$, $J=2(h_i-j)+1$. Here
$R=\{l_1\geq l_2\geq  \ldots \geq 0\}$, so that $1\leq i\leq l_j$, $1\leq j\leq h_i$
and the transposed diagram is
$R'=\{h_1\geq h_2\geq\ldots \geq 0\}$. We also define
\be
Z_{I|J}^{(s)}(A|q)=Z_{I+s|J-s}(A|q)
\ee

\item
The $Z$-factor $Z_{I|J}$ does not contribute to the Alexander polynomial
whenever $I=0$ or $J=0$.

\item
Contributions of products of the $Z$-factors corresponding to subsets with $k$ boxes
to the Alexander polynomial actually vanish as $\epsilon^{2k}$ as $\epsilon = q-q^{-1} \rightarrow 0$.

\item
The $Z$-factor $Z_{I|J}$ does not contribute to the Jones polynomial
whenever $J=2$.
Jones polynomials vanish for $R$ with three or more lines because of the
unknot factor $\SS_R(A|q)$.
Without this factor the HOMFLY polynomial does not vanish at $A=q^2$ for any $R$, while the
HOMFLY polynomial with two lines at $A=q^2$ is equal to that in the symmetric (one-line) representation $[l_1-l_2]$.

\item
The answer for the HOMFLY polynomial
is a sum over {\it all} subsets
of boxes from $R$.

\item
Coefficients in this sum are $1+O(\epsilon^2)$, the $\epsilon^2$-corrections being presumably present
only for the {\it disconnected} subsets of $R$.

\item
Each box contributes its own $Z$-factor, but the arguments are shifted
depending on the position of the box in the original diagram and in the given collection.
However, the problem is to specify the shifts.

\item
The superpolynomials are obtained by replacing the $Z$-factors by
${\mathfrak Z}_{I|J}(A) = \{Aq^I\}\{At^{-J}\}$. Moreover, it is convenient to introduce
the doubly shifted ${\mathfrak Z}$-factors
\be
{\mathfrak Z}_{I|J}^{(s|\sigma)}(A) =
{\mathfrak Z}_{I|J}(q^st^{-\sigma}A)=\{Aq^{I+s}t^{-\sigma}\}\{Aq^st^{-J-\sigma}\}
\ee
(hence, one should distinguish between $q$ and $t$ shifts),
and each such factor has positivity property in the boldface variables\footnote{In \cite{GS}
and later papers \cite{ind,Fuji} other variables were used:
\be
\tilde{\bf q}=q^2,\ \ \ \ \ \ \ \tilde{\bf t}=-{t\over q},\ \ \ \ \ \ \ \ \ \ \
\tilde{\bf a}={A^2q^3\over t^3}
\label{newnewv}
\ee
.}
\be
t={\bf q},\ \ \ \ \ \ \ q=-{\bf q}{\bf t},\ \ \ \ \ \ \ \ \ \ \
A = {\bf a}\sqrt{-{\bf t}}
\label{newv}
\ee
by itself, whenever $I$ is odd:
\be
{\mathfrak Z}_{I|J}^{(s|\sigma)}({\bf a},{\bf q},{\bf t}) =(-1)^{I+1}{
\left({\bf a}^2{\bf t}^{2I+2s+1}{\bf q}^{2(I+s-\sigma)}+1
\right)
\left({\bf a}^2{\bf t}^{2s+1}{\bf q}^{2(s-J-\sigma)}+1
\right)\over {\bf a}^2{\bf t}^{I+2s+1}{\bf q}^{I-J+2(s-\sigma)}}
\ee
Otherwise, the positive is the factor $-{\mathfrak Z}_{I|J}^{(s|\sigma)}(A)$,
i.e. the product of two such $Z$-factors.

\item
The answer for the superpolynomial for the transposed representation $R'$
is obtained by the change $(A,q,t) \longrightarrow (A,-t^{-1},-q^{-1})$.
For HOMFLY the transformation is just $q\rightarrow - q^{-1}$.
\end{itemize}

\section{Speculations about representation $R=\a$: from special,
Alexander and Jones to HOMFLY}

Jones polynomial for $\a$ is easily available: for the $SU(2)$ group
it is undistinguishable from $\f$, i.e.
\be
{\cal J}_{[21]}^{4_1}(q) = {\cal J}_{[1]}^{4_1}(q) = q^4-q^2+1-q^{-2}+q^{-4}
\ee
The second distinguished case where the answer is immediately known is
the {\it special} polynomial \cite{DMMSS} (the reduction of the HOMFLY polynomial to $q=1$),
which is always \cite{DMMSS,Zhu} expressed through $\sigma_{[1]}$:
\be
\sigma^{4_1}_{[21]}(A) = \Big(\sigma^{4_1}_{[1]}(A)\Big)^3=\Big(A^2-1+A^{-2}\Big)^3
\label{sperel}
\ee
Similarly, \cite{IMMMfe} the Alexander polynomial is immediately known from a mysterious
"dual" of (\ref{sperel}), valid only for the hook diagrams,
but $\a$ belongs to this class:
\be
\AL_{[21]}^{4_1}(q) = \AL_{[1]}^{4_1}(q^3) = -q^6+3-q^{-6}
\label{alrel}
\ee
In both (\ref{sperel}) and (\ref{alrel}) the degree 3 comes from $3=|R|=|\hbox{\a}|$.

Thus, one needs an expression built by the rules of s.\ref{gentwi} and
satisfying these constraints.

Representation $R=\a$ is still not sufficiently general, because it is
a hook diagram, still a verification problem exists already here.
Since in this case the transposed representation $R'=R$, the HOMFLY polynomial should be symmetric
under the simultaneous reflection $Z_{I|J} \rightarrow Z_{J|I}$
of all the $Z$-factors.
Also only single-box sets should contribute when $A=1$ and $A=q^2$,
in order to reproduce the Alexander and Jones polynomials, i.e.
all the 2- and 3-box sets should contain at least one
$Z_{\cdot|0}$ or $Z_{0|\cdot}$ and at least one $Z_{\cdot|2}$.
These requirements severely restrict the possible answer:
\be
\frac{{\cal H}_{[21]}(A|\,q)}{\SS_{[21]}(A|\,q)} \ \stackrel{?}{=} \
1 + \Big(Z_{3|3}+Z_{2|0}+Z_{0|2}\Big)
+ \Big(Z_{4|2}Z_{2|0} + Z_{2|4}Z_{0|2}  +  \alpha Z_{2|0}Z_{0|2}\Big)
+ Z_{3|3}Z_{2|0}Z_{0|2}
\ee
where $\SS_Q$ is the HOMFLY polynomial of the unknot in representation $Q$.
It is {\it a priori} unclear if $\alpha$ is equal to zero or not, because this term
corresponds to a disconnected subset of the Young diagram $\a$ (which is not a Young sub-diagram).

Clearly, the Alexander polynomial gets contribution from the single term:
\be
\AL_{[21]}^{4_1}(q) = 1 + Z_{3|3}(A|\,q)\Big|_{A=1} = 1 - \{q^3\}^2 =
\AL_{[1]}^{4_1}(q^3)
\ee
Similarly, the Jones polynomial is
\be
\!\!\!\!\!\! \!\!\!
{\cal J}_{[21]}^{4_1}(q) = 1 + Z_{3|3}(A|\,q) + Z_{2|0}(A|\,q)\Big|_{A=q^2} =
1 + \big(-[5]+[2]\cdot[4]\big)\{q\}^2 = 1 + [3]\{q\}^2 = {\cal J}_{[1]}^{4_1}(q)
\ee
or, in other words,
\be
\frac{{\cal H}_{[21]}(A|\,q)}{\SS_{[21]}(A|\,q)} -
\frac{{\cal H}_{[1]}(A|\,q)}{\SS_{[1]}(A|\,q)}
= \{Aq^2\}\{Aq^{-2}\}F_{[21]}(A|\,q)
\ee
with some function $F_{[21]}(A|\,q)$.
However, both these specializations are insensitive to $\alpha$ and
can not be used to decide if $\alpha$ is unity or anything else.

Here the information about the special polynomials
and especially about corrections in $\epsilon$ to formula (\ref{sperel})
found in \cite{Sle} becomes very important:

\begin{itemize}
\item
If we want that the special polynomial is equal to the cube of the fundamental
special polynomial,
\be
\sigma_{[21]}(A) = \sigma_{[1]}(A)^3 = (A^2-1+A^{-2})^3,
\ee
then
\be
{\rm at}\ \ q=1 \ \ \ {\rm the\ coefficient} \ \ \ \alpha=1
\ee
\item
The first correction, evaluated in \cite{Sle}, implies that
\be
\alpha = 1 - \epsilon^2 + O(\epsilon^4)
\ee
\end{itemize}

\section{Exact answer for HOMFLY polynomial}

Exact evaluation with the help of the cabling method \cite{Ancab}
(it requires a $9$-strand calculation and heavily relies on the results of
\cite{AMMM}) demonstrates that this answer is exact\footnote{
One could speculate that this is because the diagram $\a$ is hook:
for $k$ hooks one would probably have a polynomial of degree $k$
in $\epsilon^2$.
}:
$\boxed{\alpha = 1-\epsilon^2}$
and
$$
\boxed{
\frac{{\cal H}_{[21]}(A|\,q)}{\SS_{[21]}(A|\,q)} \ = \
1 + \Big(Z_{3|3}+Z_{2|0}+Z_{0|2}\Big)
+ \Big(Z_{4|2}Z_{2|0} + Z_{2|4}Z_{0|2}  +  (1-\epsilon^2) Z_{2|0}Z_{0|2}\Big)
+ Z_{3|3}Z_{2|0}Z_{0|2}
}
$$
\be
= \Big(A^6+A^{-6}\Big)
- ({q}^{6}+{q}^{2}-1+{q}^{-2}+{q}^{-6})\Big(A^4+A^{-4}\Big) +\nn\\
+(q^{10}-q^8+3\,q^6-3\,q^4+5\,{q}^{2}-4 +5\,{q}^{-2}-3\,{q}^{-4}+3\,{q}^{-6}-{q}^{-8}+{q}^{-10}
)\Big(A^2+A^{-2}\Big) \nn\\
-(2\,{q}^{10}-2\,{q}^{8}+5\,{q}^{6}-6\,{q}^{4}+8\,{q}^{2}-7
+8\,{q}^{-2}-6\,{q}^{-4}+5\,{q}^{-6}-2\,{q}^{-8}+2\,{q}^{-10})
\label{main}
\ee
See the Appendix for a short description of this derivation and \cite{Ancab} for a detailed presentation.

Pictorially the answer can be represented as follows:

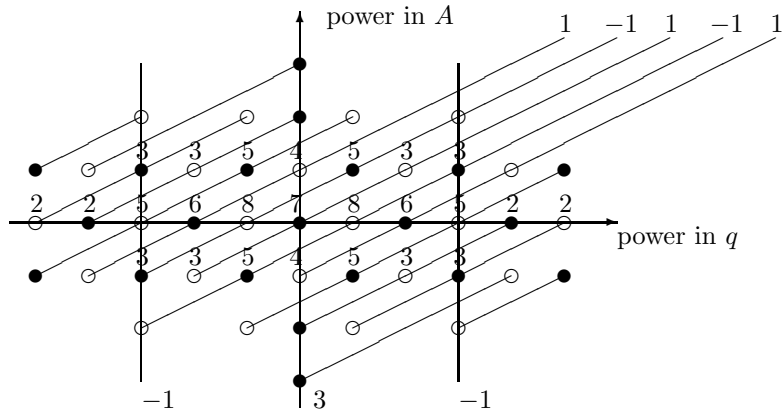
\begin{picture}(300,160)(-200,-70)
\put(-110,0){\vector(1,0){230}}
\put(0,-70){\vector(0,1){150}}
\put(0,60){\circle*{5}}  \put(0,-60){\circle*{5}}
\put(-60,40){\circle{5}} \put(-20,40){\circle{5}} \put(0,40){\circle*{5}}
\put(20,40){\circle{5}} \put(60,40){\circle{5}}
\put(-60,-40){\circle{5}} \put(-20,-40){\circle{5}} \put(0,-40){\circle*{5}}
\put(20,-40){\circle{5}} \put(60,-40){\circle{5}}
\put(-100,20){\circle*{5}} \put(-80,20){\circle{5}} \put(-60,20){\circle*{5}}\put(-62,24){\mbox{$3$}}
\put(-40,20){\circle{5}}\put(-42,24){\mbox{$3$}} \put(-20,20){\circle*{5}}\put(-22,24){\mbox{$5$}}
\put(0,20){\circle{5}}\put(-4,24){\mbox{$4$}}
\put(100,20){\circle*{5}} \put(80,20){\circle{5}} \put(60,20){\circle*{5}}\put(58,24){\mbox{$3$}}
\put(40,20){\circle{5}}\put(38,24){\mbox{$3$}} \put(20,20){\circle*{5}}\put(18,24){\mbox{$5$}}
\put(-100,-20){\circle*{5}} \put(-80,-20){\circle{5}} \put(-60,-20){\circle*{5}}\put(-62,-16){\mbox{$3$}}
\put(-40,-20){\circle{5}}\put(-42,-16){\mbox{$3$}} \put(-20,-20){\circle*{5}}\put(-22,-16){\mbox{$5$}}
\put(0,-20){\circle{5}}\put(-4,-16){\mbox{$4$}}
\put(100,-20){\circle*{5}} \put(80,-20){\circle{5}} \put(60,-20){\circle*{5}}\put(58,-16){\mbox{$3$}}
\put(40,-20){\circle{5}}\put(38,-16){\mbox{$3$}} \put(20,-20){\circle*{5}}\put(18,-16){\mbox{$5$}}
\put(-100,0){\circle{5}}\put(-102,4){\mbox{$2$}} \put(-80,0){\circle*{5}}\put(-82,4){\mbox{$2$}}
\put(-60,0){\circle{5}}\put(-62,4){\mbox{$5$}} \put(-40,0){\circle*{5}}\put(-42,4){\mbox{$6$}}
\put(-20,0){\circle{5}}\put(-22,4){\mbox{$8$}} \put(0,0){\circle*{5}}\put(-4,4){\mbox{$7$}}
\put(100,0){\circle{5}}\put(98,4){\mbox{$2$}} \put(80,0){\circle*{5}}\put(78,4){\mbox{$2$}}
\put(60,0){\circle{5}}\put(58,4){\mbox{$5$}} \put(40,0){\circle*{5}}\put(38,4){\mbox{$6$}}
\put(20,0){\circle{5}}\put(18,4){\mbox{$8$}}
\put(10,75){\mbox{power in $A$}}
\put(120,-8){\mbox{power in $q$}}
\put(-100,20){\line(2,1){40}}
\put(-80,20){\line(2,1){80}}
\put(-100,0){\line(2,1){80}}
\put(-80,0){\line(2,1){80}}
\put(-100,-20){\line(2,1){120}}
\put(-80,-20){\line(2,1){180}}
\put(-60,-20){\line(2,1){180}}
\put(-40,-20){\line(2,1){180}}
\put(-60,-40){\line(2,1){220}}
\put(0,-20){\line(2,1){180}}
\put(-20,-40){\line(2,1){120}}
\put(0,-40){\line(2,1){80}}
\put(20,-40){\line(2,1){80}}
\put(0,-60){\line(2,1){80}}
\put(60,-40){\line(2,1){40}}
\put(98,72){\mbox{$1$}}
\put(115,72){\mbox{$-1$}}
\put(138,72){\mbox{$1$}}
\put(155,72){\mbox{$-1$}}
\put(178,72){\mbox{$1$}}
\put(-60,-60){\line(0,1){120}}
\put(-60,-70){\mbox{$-1$}}
\put(60,-70){\mbox{$-1$}}
\put(60,-60){\line(0,1){120}}
\put(5,-70){\mbox{$3$}}
\end{picture}

\bigskip

Vertical and horizontal axes label powers of $A$ and $q$ respectively,
black and white circles denote plus and minus signs, non-unity
multiplicities are explicitly written over the corresponding vertices
of the Newton polygon.
The lines correspond to the case of $A=q^{-2}$: summing up algebraically the
multiplicities along each line one should get the same answer as for
the fundamental representation: $\H_{[21]}(A=q^2) = \H_{[1]}(A=q^2)=q^{-4}-q^{-2}+1-q^2+q^4$.
The symmetric set of lines (not shown in the picture) describe in the same
way the requirement $\H_{[21]}(A=q^{-2}) = \H_{[1]}(A=q^{-2})$. Similarly, the vertical lines (of which
we show only the three with non-vanishing sums) describe the specialization $A=1$:
$\H_{[21]}(A=1)=3-q^3-q^{-3}$.

\section{Next steps}

After reaching "a critical point" (\ref{main})
one can go into several directions.

One option is to go "down" and test various ideas one could have
about the colored HOMFLY polynomials and their calculations,
we continue doing this elsewhere.

Another option is to go further up in at least three directions.

\begin{itemize}
\item
One can extend (\ref{main}) to other knots:
the Appendix lists more results from \cite{Ancab}
for the 3-strand examples.
\item
One can try higher representations for the figure eight knot;
we discuss this very briefly in section \ref{hire}.
\item
One can look at the superpolynomial, see s.\ref{super}.
\end{itemize}

\section{On generic hook diagram $R=[b,1^{a-1}]$
\label{hire}}

\subsection{Hook diagram and the answer }

Generic hook diagram $[r,1^{s-1}]$, has $r$ columns and $s$ lines:

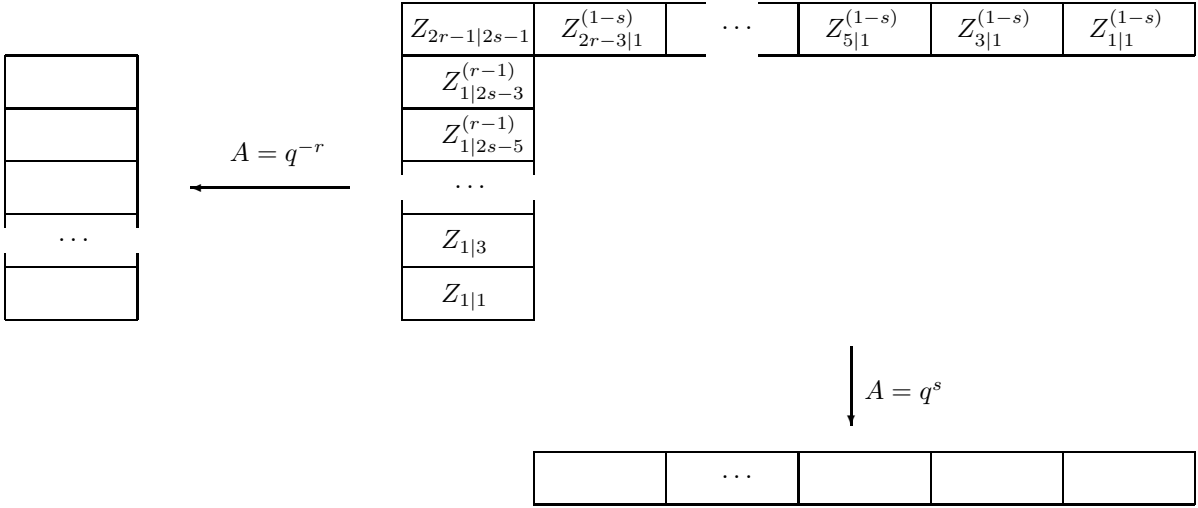
\begin{picture}(200,220)(-150,-90)
\put(0,0){\line(0,1){45}}
\put(50,0){\line(0,1){45}}
\put(0,55){\line(0,1){65}}
\put(50,55){\line(0,1){65}}
\put(0,100){\line(1,0){115}}
\put(0,120){\line(1,0){115}}
\put(135,100){\line(1,0){165}}
\put(135,120){\line(1,0){165}}
\put(0,40){\line(1,0){50}}
\put(0,60){\line(1,0){50}}
\put(0,80){\line(1,0){50}}
\put(0,100){\line(1,0){50}}
\put(0,120){\line(1,0){50}}
\put(3,107){\makebox{{$Z_{2r-1|2s-1}$}}}
\put(15,87){\makebox{{$Z_{1|2s-3}^{(r-1)}$}}}
\put(15,67){\makebox{{$Z_{1|2s-5}^{(r-1)}$}}}
\put(20,50){\makebox{$\ldots$}}
\put(15,27){\makebox{{$Z_{1|3}$}}}
\put(15,7){\makebox{{$Z_{1|1}$}}}
\put(260,107){\makebox{{$Z_{1|1}^{(1-s)}$}}}
\put(210,107){\makebox{{$Z_{3|1}^{(1-s)}$}}}
\put(160,107){\makebox{{$Z_{5|1}^{(1-s)}$}}}
\put(121,110){\makebox{$\ldots$}}
\put(60,107){\makebox{{$Z_{2r-3|1}^{(1-s)}$}}}
\put(100,100){\line(0,1){20}}
\put(150,100){\line(0,1){20}}
\put(200,100){\line(0,1){20}}
\put(250,100){\line(0,1){20}}
\put(300,100){\line(0,1){20}}
\put(0,0){\line(1,0){50}}
\put(0,20){\line(1,0){50}}
\put(-150,0){\line(0,1){25}}
\put(-100,0){\line(0,1){25}}
\put(-150,35){\line(0,1){65}}
\put(-100,35){\line(0,1){65}}
\put(-150,0){\line(1,0){50}}
\put(-150,20){\line(1,0){50}}
\put(-150,40){\line(1,0){50}}
\put(-150,60){\line(1,0){50}}
\put(-150,80){\line(1,0){50}}
\put(-150,100){\line(1,0){50}}
\put(-130,30){\makebox{$\ldots$}}
\put(-20,50){\vector(-1,0){60}}
\put(-65,60){\makebox{$A=q^{-r}$}}
\put(50,-70){\line(1,0){115}}
\put(50,-50){\line(1,0){115}}
\put(135,-70){\line(1,0){165}}
\put(135,-50){\line(1,0){165}}
\put(50,-70){\line(0,1){20}}
\put(100,-70){\line(0,1){20}}
\put(121,-60){\makebox{$\ldots$}}
\put(150,-70){\line(0,1){20}}
\put(200,-70){\line(0,1){20}}
\put(250,-70){\line(0,1){20}}
\put(300,-70){\line(0,1){20}}
\put(170,-10){\vector(0,-1){30}}
\put(175,-30){\makebox{$A=q^{s}$}}
\end{picture}

\noindent
The relevant set of the $Z$-factors for this hook diagram  is:
\be
Z_{1|1},Z_{3|1},Z_{5|1},\ldots,Z_{2r-3|1}, \ \ \ \ \
Z_{2r-1|2s-1},\ \ \ \ \
Z_{1|2s-3},Z_{1|2s-5},\ldots, Z_{1|3},Z_{1|1}
\ee
for the line, corner and column respectively.

In the $Z$-linear terms all the line-factors are shifted by $1-s$
and all the column-factors by $r-1$:
$$
\frac{\HH_{hook}}{^{^*}S_{hook}}
= 1 + \Big(Z_{1|1}^{(1-s)} + Z_{3|1}^{(1-s)}
+ \ldots
+ Z_{2r-3|1}^{(1-s)}\Big) + Z_{2r-1|2s-1}  +
\Big(Z_{1|2s-3}^{(r-1)} + \ldots
+ Z_{1|3}^{(r-1)} + Z_{1|1}^{(r-1)}\Big) + O(Z^2) =
$$
\vspace{-0.5cm}
\be
= \left\{\Big(Z_{2-s|s}+Z_{4-s|s} + \ldots + Z_{2r-2-s|s}\Big)
+ Z_{2r-1|2s-1} + \Big(Z_{r|2s-2-r} + \ldots + Z_{r|4-r} + Z_{r|2-r}\Big)\right\}
+ O(Z^2)
\label{hooklin}
\ee

\subsection{Validation at the level of $Z$-linear terms}

There are five things to check about this formula.

$\bullet$ First of all, at $q=1$ one should get a special polynomial
with the factorization property:
\be
\frac{\HH_{[r,1^{s-1}]}}{^{^*}S_{[r,1^{s-1}]}}(q=1|A) =
\left(\frac{\HH_{_\Box}}{^{^*}S_{\Box}}(q=1|A)\right)^{r+s-1}
\ee
This is built in the general construction for the figure eight knot,
since at $q=1$ all the $Z$-factors coincide, $\epsilon$-corrections
are absent, and the weighted sum over all subsets of boxes in the
Young diagram  is immediately equal to $(1+Z)^{|R|}$.

$\bullet$
at $A=1$ one should get the Alexander polynomial
\be
\AL_{hook}(q) = \AL_{_\Box}(q^{r+s-1}) = 1 - [r+s-1]^2\epsilon^2
\ee
and this should hold at the level of linear terms.
It is reproduced by (\ref{hooklin}), because
\be
-[s]\cdot\Big([2-s]+[4-s]+\ldots+[2(r-1)-s]\Big) -[2r-1]\cdot[2s-1] + \nn \\
+ [r]\cdot\Big([r-2]+[r-4]+\ldots + [r-2(s-1)]\Big) \ \ =  \ \  -[s+r-1]^2
\ee

$\bullet$ Third, at $A=q^s$ the answer should coincide with the
HOMFLY polynomial for symmetric representation $[r-1]$:
\be
\frac{\HH_{[r,1^{s-1}]}}{^{^*}S_{[r,1^{s-1}]}}(A=q^s)
= \frac{\HH_{[r-1]}}{^{^*}S_{[r-1]}}(A=q^s) = \nn \\
= \left.\Big(1 + [r-1]\cdot\{Aq^{r-1}\}\{A/q\} + O(Z^2)\Big)\right|_{A=q^s}
= 1+[r-1]\cdot[s-1]\cdot[r+s-1]\epsilon^2 + O(\epsilon^4)
\ee
Eq.(\ref{hooklin}) satisfies this because
\be
0 +[s+2r-1]\cdot[s+1-2s]  + [s+r]\Big([s+r-2]+[s+r-4]+\ldots + [s+r-2(s-1)]\Big)=
\nn \\
=  [r-1]\cdot[s-1]\cdot[r+s-1]
\ee

$\bullet$ Fourth, at $A=q^{-r}$ the answer should coincide with the
HOMFLY polynomial for the antisymmetric representation $[1^{s-1}]$:
\be
\frac{\HH_{[r,1^{s-1}]}}{^{^*}S_{[r,1^{s-1}]}}(A=q^{-r})
= \frac{\HH_{[1^{s-1}]}}{^{^*}S_{[r-1]}}(A=q^{-r})
= \nn \\
= \left.\Big(1 + [s-1]\cdot\{Aq^{1-s}\}\{Aq\} + O(Z^2)\Big)\right|_{A=q^{-r}}
= 1 + [r-1]\cdot[s-1]\cdot[r+s-1]\epsilon^2 + O(\epsilon^4)
\ee
This time eq.(\ref{hooklin}) satisfies this because
\be
[-r-s]\cdot\Big([-r+2-s]+[-r+4-s]+\ldots+[-r+2(r-1)-s]\Big) +[-r+2r-1]\cdot[-r+1-2s] + 0 =
\nn \\
=   [r-1]\cdot[s-1]\cdot[r+s-1]
\ee

$\bullet$
The fifth
observation is that the sum of left indices is related to $\nu_{[r,1^{s-1}]} = \frac{s(s-1)}{2}$,
while that of the right indices to $\nu_{[s,1^{r-1}]} = \frac{r(r-1)}{2}$ for the transposed diagram.
More precisely,
\be
1+3+5+\ldots+(2r-3)+(2r-1) + 1 + \ldots + 1 + 1 + 1 = r^2+s-1  =
r+s-1 + 2\,\frac{r(r-1)}{2} = |R| + \nu_{R'}, \nn \\
1 + 1 + 1 + \ldots + 1 + (2s-1) + (2s-3) + \ldots + 5 + 3 + 1 = s^2+r-1 = r+s-1 + 2\,\frac{r(r-1)}{2}
= |R| + \nu_{R}
\ee
(note that the shifts do not contribute to the sum: $2(r-1)\cdot(1-s) + 2(s-1)\cdot (r-1)=0$).
As we shall see, this fact is important for superpolynomial studies, see eq.(\ref{ant2}).

\subsection{Higher-order terms in $Z$, high degree $\epsilon$-corrections
and other generalizations}

If $s=2$
the higher order terms should vanish when $A=q^{-r}$ and $A=1$,
since both the Alexander and fundamental HOMFLY polynomials get only
contributions from the $Z$-linear terms, while
for $A=q^2$ one should obtain the Jones polynomial (i.e. HOMFLY in the symmetric
representation $[r-1]$ at $A=q^2$ \cite{IMMMfe}), and $Z$-quadratic and higher terms
do contribute (for $r\geq 3$), but we know
what these contributions
are comparing them with the $Z$-factor expansion of \cite{IMMMfe}.
When $s>2$ higher order terms are also present when $A=q^{-r}$,
but again we know \cite{IMMMfe} the explicit expression for the
antisymmetric representations.

Assuming that the only parameters that depend
on  $r$ and $s$ are the integer-valued shifts,
one can adjust them, first looking at the expansion in powers of $\epsilon$
order-by-order in $Z$, and
solving simple linear equations for the coefficients
of polynomials in $r$ and $s$ (which are restrictive,
because rarely have integer-valued solutions).
After that one can check that they are true for
arbitrary $q$.

The $\epsilon^2$-corrections to the integer valued
coefficients can be further controlled by the
$\epsilon$-expansion of \cite{Sle}.
This is rather a constructive procedure, which will be
described in more detail elsewhere. It is important,
because at the moment it looks hardly possible
to extend the calculation of \cite{Ancab}, relying on the
cabling method to higher representations (unless
powerful computer is used to multiply huge matrices:
if it is available, the calculation is straightforward).
The colored eigenvalue approach of \cite{IMMM3}
should be computationally easier, but still needs
to be better understood and developed.
In these circumstances the $Z$-expansion approach
of \cite{IMMMfe} can be competitive (unfortunately, at the moment
it is restricted to the figure eight and other twist
knots \cite{ind,twist}).

The next subject to discuss are superpolynomials.
As already explained in \cite{IMMMfe} and confirmed in
\cite{ind,Fuji} the $Z$-expansion makes the $t$-deformation
almost algorithmic, modulo some open questions and
controversies about the theory and the very notion of
the superpolynomial itself.
We now proceed to a brief discussion of this subject.

\section{On superpolynomial for $R=\a$
\label{super}}

The story of colored superpolynomials is today one of the most interesting
and puzzling.
Even in the Khovanov-Rozansky approach there is still no unambiguous definition
and reliable results, nothing to say about the clear definition of
colored superpolynomial itself.

The case of the figure eight knot \cite{IMMMfe},
and partly of the other twist knots \cite{ind,Fuji,twist}
look a lucky exception, because the $t$-deformation
in the $Z$-factor representation
(closely related to the DGR differentials)
was "obvious" and straightforward for symmetric and duality-related
antisymmetric representations.
As we shall see, however, in other representations the idyll
is still to be found: already in the simplest maximally symmetric
case of $P_{[21]}^{4_1}$ there are ambiguities, at least in
the naive approach.

\subsection{Requirements}

What one needs is a $t$-deformation of (\ref{main}) with the following
properties:

\begin{itemize}
\item[{\bf (A)}] It reproduces the HOMFLY polynomial (\ref{main}) at $q=t$,
\be
P_{R}^{{\cal K}}(A|\,q,t=q) =H_{R}^{{\cal K}}(A|\,q)
\ee
Putting further $q=t=1$ one obtains the factorization property
(\ref{sperel}) of the special polynomials.

\item[{\bf (B)}] All coefficients in front of all monomials
$(-q)^kt^l(-A^2)^m = {\bf q}^{k+l}{\bf a}^{2m}{\bf t}^{k+m}$
are positive integers.

\item[{\bf (C)}] There is a symmetry (duality \cite{DMMSS} or mirror \cite{GS}),
$(A,t,q) \leftrightarrow (A,-q^{-1},-t^{-1})$ or $\Big({\bf a},{\bf q},{\bf t}\Big)
\leftrightarrow \Big({\bf a},({\bf q}{\bf t})^{-1},{\bf t}\Big)$:
\be
P_{R}^{{\cal K}}(A|\,q,t) = P_{R'}^{{\cal K}}\Big(A\Big|-t^{-1},-q^{-1}\Big)
\label{sd}
\ee
where $R'$ is the transposed Young diagram.
$R = \a$, like $R=\f$, is a self-dual case.

\item[{\bf (D)}] For $A=t^2$ and $A=q^{-2}$ the answer in representation $\a$
coincides with that in the fundamental representation $\f$,
\be
P_{[21]}^{{\cal K}}(A=t^2) = P_{[1]}^{{\cal K}}(A=t^2), \nn \\
P_{[21]}^{{\cal K}}(A=q^{-2}) = P_{[1]}^{{\cal K}}(A=q^{-2})
\label{Jo}
\ee
These two equations coincide, once (\ref{sd}) is true.
More generally, for a hook diagram $R=[r,1^{s-1}]$
\be
P_{[r,1^{s-1}]}^{{\cal K}}(A=t^s) = P_{[r-1]}^{{\cal K}}(A=t^s), \nn \\
P_{[r,1^{s-1}]}^{{\cal K}}(A=q^{-r}) = P_{[1^{s-1}]}^{{\cal K}}(A=q^{-r})
\ee
In boldface variables our conditions $A=t^s$ and $A=q^{-r}$
turn into the DGR-differential conditions
$\ {\bf a}^2{\bf t} +{\bf q}^{2s} = 0\ $ and
$\ {\bf a}^2{\bf t}^{2r}{\bf t}^{2r+1} + 1 =0$.

\item[{\bf (E)}] For $A=t/q\ $  (i.e. $\ {\bf a}^2{\bf t}^3+1=0$)\
the answer in representation $\a$
(Heegard-Floer polynomial)
satisfies
\be
P_{[21]}^{{\cal K}}(A={t\over q}\,\big|\, q,t) = P_{[1]}^{{\cal K}}(A={t\over q}\,\big|\, q^2t,t^2q)
\ee
More generally
\be
\boxed{
P_{[r,1^{s-1}]}^{{\cal K}}(A={t\over q}\,\big|\, q,t) = P_{[1]}^{{\cal K}}(A={t\over q}\,\big|\, q^rt^{s-1},t^sq^{r-1})
}
\ee
This is generalization of the property (\ref{alrel})
for the Alexander polynomial.

The Khovanov-Rozansky and Heegard-Floer polynomials are obtained after throwing away the terms
of the original superpolynomials canceling with each other at $A=t^N$ by the other substitutions
$A=t^N\sqrt{q/t}$ (i.e. ${\bf a}={\bf q}^N$) and
$A=\sqrt{t/q}$ (i.e. ${\bf a}={\bf t}^{-1}$) respectively
\cite{DGR}.

\item[{\bf (F)}] The first deviation from the special polynomial
for $q=e^\hbar$, $t=e^{\bar\hbar}$
is given by \cite{Ant2}
\be
P_R^{{\cal K}}(A|e^\hbar,e^{\bar\hbar})
= \left(P_{[1]}^{{\cal K}}(A|e^\hbar,e^{\bar\hbar})\right)^{|R|}
+ (\hbar\nu_{R'} - \bar\hbar\nu_{R})\sigma_1^{|R|-2}(A)\sigma_2(A)
+ O(\hbar^2,\bar\hbar^2,\hbar\bar\hbar)
\label{ant2}
\ee
For the figure eight knot ${\cal K}=4_1$
\be
P_{[1]}^{4_1} = 1 + {\cal Z}_{1|1} = 1+\{Aq\}\{A/t\},
\ \ \ \ \ \sigma_1(A) = 1+\{A\}^2, \ \ \ \ \sigma_2\{A\} = 2\{A^2\}\Big(1+2\{A\}^2\Big)
\ee
and, since $\nu_{21}=1$, eq.(\ref{ant2}) claims that
\be
P_{[21]}^{4_1} =
1+(\hbar-\bar\hbar)\{A^2\}\big(3\sigma_1^2+\sigma_1\sigma_2\big) + O(\hbar^2,\bar\hbar^2,\hbar\bar\hbar)=
1+(\hbar-\bar\hbar)\{A^2\}\Big(5+12\{A\}^2+7\{A\}^4\Big)
+\ldots
\label{anti}
\ee
Eq.(\ref{ant2}) is conjectured on the base of three arguments:
the factorization property of "special superpolynomials" in the symmetric
representations \cite{Ant1}, the evolution hypothesis of \cite{DMMSS}
and the symmetric group character expansion of the HOMFLY polynomials of \cite{MMpols,Sle}.
There is actually no way to test the formula itself, since there is no yet
a single example known of a non-trivially colored superpolynomial,
however, in the symmetric representation (\ref{anti}) is a particular case of much more general
factorization hypothesis (which by now is violated by a single example
of $P_{[2]}^{9_{42}}$ in \cite{GS}, but is certainly true
for the twist knots \cite{ind} including $4_1$).
\end{itemize}

It would be very nice to apply the other criteria of \cite{DGR,GS,MMpols}
to study of $P_{[21]}^{4_1}$, but this remains to be done:
here we consider only the above six items which are unambiguously formulated.

\subsection{Modifications within the $Z$-factor expansion}

Following \cite{IMMMfe}, these criteria are easy to study by a simple deformation
of $Z$-factors.
Since the Young diagram $\a$ spreads beyond one column and one line,
we need two kinds of shifts: in the horizontal and
vertical directions so that the relevant ${\mathfrak Z}$-factor is going to be
\be
{\mathfrak Z}_{I|J}^{(s|\sigma)}(A) =
{\mathfrak Z}_{I|J}(q^st^{-\sigma}A)=\{Aq^{I+s}t^{-\sigma}\}\{Aq^st^{-J-\sigma}\}
= \nn \\
=(-)^{I+1}\frac{({\bf a}^2{\bf q}^{2(I+s)}{\bf t}^{2I+2s+1}+{\bf q}^{2\sigma})
({\bf a}^2{\bf q}^{2s}{\bf t}^{2s+1}+{\bf q}^{2(J+\sigma)})}
{{\bf a}^2{\bf q}^{I+J+2s+ 2\sigma}{\bf t}^{I+2s+1}}
\ee
Following \cite{IMMMfe}, one should just substitute the $Z$-factors in (\ref{main})
by some ${\mathfrak Z}$-factors,
trying to satisfy our criteria and taking into account that
\begin{itemize}
\item[(A)] implies that $Z_{i|j}$ goes into ${\mathfrak Z}_{I|J}^{(s|\sigma)}$
with $I+s-\sigma=i$ and $J+\sigma-s =j$.
\item[(B):] ${\mathfrak Z}$-factor has a positivity property in bold variables whenever $I$ is odd.
For even $I$ the positivity property is possessed by $-{\mathfrak Z}$ so that
a product of two ${\mathfrak Z}$-factors with even $I$ is also acceptable.
\item[(C):] Each ${\mathfrak Z}_{I|J}^{(s|\sigma)}$ is accompanied by
${\mathfrak Z}_{J|I}^{(\sigma|s)}$.
\item[(D+E):] Terms quadratic and cubic in $Z$-factors should disappear
for $A=t^2$, $A=q^{-2}$ and $A=t/q$,
i.e. each item contains a product of three factors
$\{Aq^2\}\{Aq/t\}\{A/t^2\}$.
For (D) it would be enough to have
${\mathfrak Z}$-factors with
a pair of $s$ and $J$ being $s=0,\ J+\sigma=2$,
a pair of $\sigma$ and $I$ being $\sigma=0,\ I+s=2$
and a pair being either $I+s=1,\ \sigma=1$
or $s=1,\ J+\sigma=1$.
\item[(F):] The contribution of each ${\mathfrak Z}_{I|J}^{(s|\sigma)}$ is
$\{A\}^2 + \{A^2\}\Big( ( I+2s  )\hbar  - ( J+2\sigma )\bar\hbar \Big) + \ldots$
\end{itemize}

\subsection{A possible answer}

It is straightforward to write down an
expression, satisfying all the requirements:
\be
\boxed{
\frac{\*{\cal P}_{[21]}(A|\,q,t)}{^*\!M_{[21]}(A|\,q,t)}  \stackrel{?}{=}
1 + \Big({\mathfrak Z}_{3|3}^{(-1|-1)}\!\!+{\mathfrak Z}_{1|1}^{(1|0)}+{\mathfrak Z}_{1|1}^{(0|1)}\Big)
+ \Big(
{\mathfrak Z}_{3|3}^{(0|-1)}{\mathfrak Z}_{1|1}^{(1|0)} + {\mathfrak Z}_{3|3}^{(-1|0)}
{\mathfrak Z}_{1|1}^{(0|1)}
+ \alpha \underline{{\mathfrak Z}_{1|1}^{(1|0)}{\mathfrak Z}_{1|1}^{(0|1)}}\Big)
+ {\mathfrak Z}_{3|3}\underline{{\mathfrak Z}_{1|1}^{(1|0)}{\mathfrak Z}_{1|1}^{(0|1)}}
}
\label{sumain}
\ee
Note that many ambiguities, for example the choice between
${\mathfrak Z}_{2|0}{\mathfrak Z}_{0|2} = \{Aq^2\}\{A\}^2\{At^{-2}\}$
and
${\mathfrak Z}_{1|1}^{(1|0)}{\mathfrak Z}_{1|1}^{(0|1)} = \{Aq^2\}\{Aq/t\}^2\{At^{-2}\}$
in the underlined products are unsensitive to all the criteria except for (\ref{anti}):
thus, the conjecture (\ref{sumain}) heavily relies on it.

The somewhat unexpected negative shifts in the first linear term
${\mathfrak Z}_{3|3}^{(-1|-1)}$ seem absolutely necessary for the reduction property
(\ref{Jo}). Amusingly, after that, {\it both} the positivity and (\ref{anti}) dictate
the choice ${\mathfrak Z}_{1|1}^{(1|0)}+{\mathfrak Z}_{1|1}^{(0|1)}$ instead of
${\mathfrak Z}_{2|0}+{\mathfrak Z}_{0|2}$ for the other linear terms.

\bigskip

The only thing which remains to be guessed is $\alpha$.
Positivity (together with the duality/mirror symmetry) seems to fix it to be
\be
\boxed{
\alpha = 1 - (q-t^{-1})(t-q^{-1}) =
1 + \Big({\bf q}{\bf t} + {\bf q}^{-1}\Big)\Big({\bf q} + {\bf q}^{-1}{\bf t}^{-1}\Big)
}
\ee
if one requires it to be unity at the self-dual point $qt=1$, see s.\ref{selfdual}.

\subsection{Numerology of the answer: can there be a {\it minimal} superpolynomial?}

In the theory of superpolynomials a big issue is the study of "minimality"
properties: the question is if some terms can be thrown away from the
superpolynomial without violating the properties (A-F) or, at least, (A-E).
The simplest example is provided at the Jones level (see eq.(22) of \cite{DM2}):
the product of the ordinary {\it reduced} superpolynomial and
the MacDonald dimension can be further "diminished" to give a smaller
{\it unreduced} superpolynomial:
\be
({\bf t}+{\bf t}^{-1})\ ^{\rm red}\!{\cal J}_{[1]}^{3_1}
= ({\bf t}+{\bf t}^{-1})({\bf q}^2 + {\bf q}^6{\bf t}^2+{\bf q}^8{\bf t}^3)
= \ ^{\rm unred}\!{\cal J}_{[1]}^{3_1} +   {\bf q}^7 {\bf t}^2(1+{\bf t})
\ee
where
\be
^{\rm unred}\!{\cal J}_{[1]}^{3_1} = {\bf q} + {\bf q}^3 + {\bf q}^5 {\bf t}^2
+ {\bf q}^9 {\bf t}^3
\ee
still possesses the positivity property, while being "smaller":
it contains just $4<2\cdot 3$ items.

A natural question arises, if
formula (\ref{sumain}) provides a "minimal possible"
superpolynomial with the positivity property and what at all
is the criterium of minimality.

One could just start from putting some powers of ${\bf t}$ in front
of each term in the HOMFLY polynomial, odd or even depending on the
sign of the coefficient. This of course provides a polynomial with the
positivity property, but for an exception of a few simple knots in the
fundamental representation,
it neither  satisfies the reduction properties
like (D) and (E), nor has anything to do with the Khovanov-Rozansky
polynomials.
In the generic case, correction terms proportional to $(1+{\bf t})$ should be added.

A second, more sophisticated observation could be that the
HOMFLY polynomial arises when ${\bf t}=-1$, when many
cancelations can occur: for example, already for the trefoil $3_1$
in the fundamental representation the HOMFLY polynomial has $3$ terms instead of the
"natural" number $5$ typical for all the twist knots,
due to an "accidental" cancelation:
\be
H^{3_1}_{_\Box} = 1-A^2\{Aq\}\{A/q\} = -A^4 + A^2(q^2+q^{-2})
\ee
From this point of view, the "natural" number of terms in the representation
$R$ for all the twist knots would be $5^{|R|}$ (counted with multiplicities).
Amusingly, this is indeed the case for our HOMFLY in (\ref{main}),
and the crucial role here is played by the $\epsilon^2$ correction
in $\alpha$.
Namely,
$125=1+3\cdot 4 + 3\cdot 4^2 + 4^3$ is exactly the
number of terms that the combination of $Z$-factors in (\ref{main}) would naively
have, because each $Z$-factor consists of $2^2=4$ items.
However, if $\alpha = 1$ there would be considerable cancelations:
a total of $65 = 33+32$ terms actually survive:

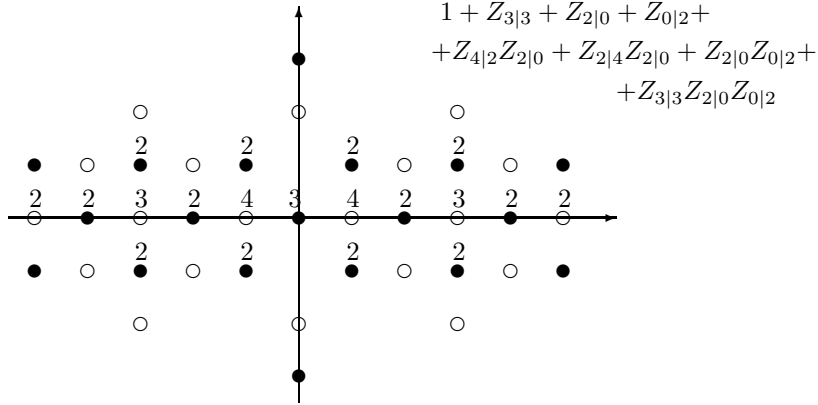
\begin{picture}(300,160)(-200,-70)
\put(-110,0){\vector(1,0){230}}
\put(0,-70){\vector(0,1){150}}
\put(0,60){\circle*{5}}  \put(0,-60){\circle*{5}}
\put(-60,40){\circle{5}}
\put(0,40){\circle{5}}
\put(60,40){\circle{5}}
\put(-60,-40){\circle{5}}
\put(0,-40){\circle{5}}
\put(60,-40){\circle{5}}
\put(-100,20){\circle*{5}} \put(-80,20){\circle{5}} \put(-60,20){\circle*{5}}\put(-62,24){\mbox{$2$}}
\put(-40,20){\circle{5}} \put(-20,20){\circle*{5}}\put(-22,24){\mbox{$2$}}
\put(100,20){\circle*{5}} \put(80,20){\circle{5}} \put(60,20){\circle*{5}}\put(58,24){\mbox{$2$}}
\put(40,20){\circle{5}} \put(20,20){\circle*{5}}\put(18,24){\mbox{$2$}}
\put(-100,-20){\circle*{5}} \put(-80,-20){\circle{5}} \put(-60,-20){\circle*{5}}\put(-62,-16){\mbox{$2$}}
\put(-40,-20){\circle{5}} \put(-20,-20){\circle*{5}}\put(-22,-16){\mbox{$2$}}
\put(100,-20){\circle*{5}} \put(80,-20){\circle{5}} \put(60,-20){\circle*{5}}\put(58,-16){\mbox{$2$}}
\put(40,-20){\circle{5}}  \put(20,-20){\circle*{5}}\put(18,-16){\mbox{$2$}}
\put(-100,0){\circle{5}}\put(-102,4){\mbox{$2$}} \put(-80,0){\circle*{5}}\put(-82,4){\mbox{$2$}}
\put(-60,0){\circle{5}}\put(-62,4){\mbox{$3$}} \put(-40,0){\circle*{5}}\put(-42,4){\mbox{$2$}}
\put(-20,0){\circle{5}}\put(-22,4){\mbox{$4$}} \put(0,0){\circle*{5}}\put(-4,4){\mbox{$3$}}
\put(100,0){\circle{5}}\put(98,4){\mbox{$2$}} \put(80,0){\circle*{5}}\put(78,4){\mbox{$2$}}
\put(60,0){\circle{5}}\put(58,4){\mbox{$3$}} \put(40,0){\circle*{5}}\put(38,4){\mbox{$2$}}
\put(20,0){\circle{5}}\put(18,4){\mbox{$4$}}
\put(50,75){\mbox{ ${1}+ {Z_{3|3}} + Z_{2|0} +  {Z_{0|2}} + $}}
\put(50,60){\mbox{$+Z_{4|2}Z_{2|0}+Z_{2|4}Z_{2|0} + Z_{2|0}Z_{0|2} +  $}}
\put(120,45){\mbox{$+ Z_{3|3}Z_{2|0}Z_{0|2}$}}
\end{picture}

\noindent
The lacking terms are added because $\alpha\neq 1$:

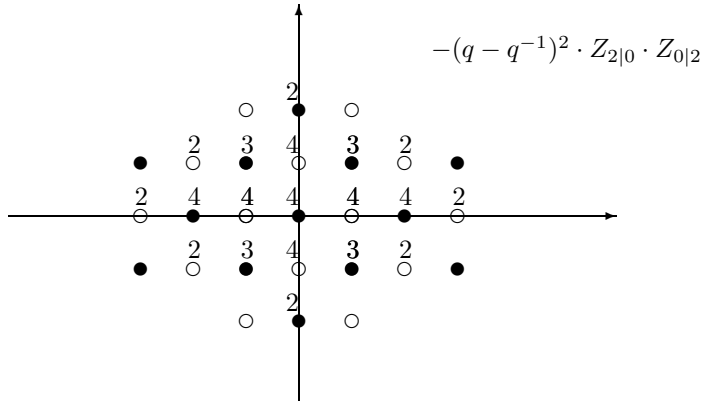
\begin{picture}(300,160)(-200,-70)
\put(-110,0){\vector(1,0){230}}
\put(0,-70){\vector(0,1){150}}
\put(0,40){\circle*{5}}\put(-5,44){\mbox{$2$}}  \put(0,-40){\circle*{5}}\put(-5,-36){\mbox{$2$}}
\put(-40,20){\circle{5}}\put(-42,24){\mbox{$2$}}
\put(0,20){\circle{5}}\put(-5,24){\mbox{$4$}}
\put(40,20){\circle{5}}\put(38,24){\mbox{$2$}}
\put(-40,-20){\circle{5}}\put(-42,-16){\mbox{$2$}}
\put(0,-20){\circle{5}}\put(-5,-16){\mbox{$4$}}
\put(40,-20){\circle{5}}\put(38,-16){\mbox{$2$}}
\put(-40,0){\circle*{5}}\put(-42,4){\mbox{$4$}}
\put(0,0){\circle*{5}}\put(-5,4){\mbox{$4$}}
\put(40,0){\circle*{5}}\put(38,4){\mbox{$4$}}
\put(20,40){\circle{5}}  \put(20,-40){\circle{5}}
\put(-20,20){\circle*{5}} \put(20,20){\circle*{5}}\put(18,24){\mbox{$3$}}
\put(60,20){\circle*{5}}
\put(-20,-20){\circle*{5}} \put(20,-20){\circle*{5}}\put(18,-16){\mbox{$3$}}
\put(60,-20){\circle*{5}}
\put(-20,0){\circle{5}}\put(-22,4){\mbox{$4$}}
\put(20,0){\circle{5}}\put(18,4){\mbox{$4$}}
\put(60,0){\circle{5}}\put(58,4){\mbox{$2$}}
\put(-20,40){\circle{5}}  \put(-20,-40){\circle{5}}
\put(-60,20){\circle*{5}} \put(-20,20){\circle*{5}}\put(-22,24){\mbox{$3$}}
\put(20,20){\circle*{5}}\put(18,24){\mbox{$3$}}
\put(-60,-20){\circle*{5}} \put(-20,-20){\circle*{5}}\put(-22,-16){\mbox{$3$}}
\put(20,-20){\circle*{5}}\put(18,-16){\mbox{$3$}}
\put(-60,0){\circle{5}}\put(-62,4){\mbox{$2$}}
\put(-20,0){\circle{5}}\put(-22,4){\mbox{$4$}}
\put(20,0){\circle{5}}\put(18,4){\mbox{$4$}}
\put(50,60){\mbox{$-(q-q^{-1})^2\cdot Z_{2|0}\cdot Z_{0|2}$}}
\end{picture}

\noindent
This provides $64=32+32$ additional terms, and there is still a cancelation
of two $A^4+A^{-4}$, what gives $65+64 - 2\cdot 2 = 125$, shown in the picture
in section 4.

At the same time our superpolynomial (\ref{sumain}) contains $189$ terms:
this number is easily calculated by putting ${\bf a}={\bf q}={\bf t}=1$.
Because of this it explicitly violates the
number-matching property
\be
\hspace{-3.4cm}{\bf (G??)}:\hspace{4.5cm}\frac{\*{\cal P}_{R}}{^*\!M_{R}} =
\left(\frac{\*{\cal P}_{[1]}}{^*\!M_{[1]}}\right)^{|R|}
\ \ \ \ \ {\rm for} \ \  {\bf a}={\bf q}={\bf t}=1 \ \ \ \  {\Large ??}
\label{numa}
\ee
which is satisfied for symmetric and antisymmetric representations
and, in general, whenever factorization property of \cite{Ant1} is correct
(it is claimed \cite{GS} to be violated even for these representations
for sufficiently complicated knots, but it certainly holds for all the
twist knots).
Amusingly, $189-125=64$ is exactly the number of terms, added by the
$\epsilon^2$ correction to $\alpha$: it has made the HOMFLY polynomial
"naturally big", but seems to make the superpolynomial too large.

Of course, looking at particular terms of the superpolynomial
expansion in powers of ${\bf a}$ and ${\bf q}$, one observes that
many coefficients contain positive contributions proportional to
$(1+{\bf t})$: if these contributions are "subtracted",
i.e. thrown away, one would get a "minimal" superpolynomial
satisfying (\ref{numa}).
However, this subtraction violates the properties
(D), (E) and (F).
One can think that (F) is not so important, still what happens
is interesting by itself: it turns out that {\it any} subtraction
{\it increases} the values of the three coefficients ($5,12,7$)
at the r.h.s. of (\ref{anti}),
a kind of a new positivity property (or minimality principle)
can be hidden here. If one wanted (\ref{anti}) to hold {\it after}
the subtraction, one should start from an expression with a lower
value of the coefficients (this can actually be done by
changing the shifts in some ${\mathfrak Z}$-factors in (\ref{sumain})).
Anyhow, violation of (D) and (E) seems unacceptable,
if one wants to preserve any relation between superpolynomials
and representation theory.
It looks like there is no way to make a subtraction and diminish
the number of terms in (\ref{sumain}), making it smaller than $189$.

The question arises, what is then the proper generalization of
(\ref{numa}) from (anti)symmetric to general representations $R$.
We try to suggest a possible answer in the next subsection.

\subsection{The second self-dual point
and the compromise between \cite{Ant1}, \cite{Ant2} and \cite{Sle}
\label{selfdual}}

In fact, there is one more shadow over (\ref{sumain}),
which {\it could} suggest that our choice of shifts might be still
modified.
It comes from consideration of the self-dual point
$qt=1$. At the other self-dual point, $q=t$ (i.e. ${\bf t}=-1$)
the superpolynomial is reduced to the HOMFLY one
and one can assume that something interesting
can happen here too.
Indeed, at $qt=1$ the coefficient $\alpha=1$
and the superpolynomial (\ref{sumain})
turns into
\be
(\ref{sumain}) \longrightarrow \ \
1+\{Aq\}^2 + 2\{Aq^2\}^2 + 3\{Aq^2\}^4 + \{Aq^3\}\{Aq^2\}^4
\ee
which is suspiciously close  to either
\be
P_{\Box}^3(qA) \ \stackrel{qt=1}{\longrightarrow}\
\Big(1+\{Aq^2\}^2\Big)^3
\ee
or to
\be
P_{\Box}(qA)P_{\Box}(A)P_{\Box}(A/t)  \ \stackrel{qt=1}{\longrightarrow}\
\Big(1+\{Aq\}^2\Big)\Big(1+\{Aq^2\}^2\Big)^2
\ee
If there was such a coincidence, this would provide
a natural generalization of the factorization properties \cite{Ant1}
\be
P_{[r]}(A) \ \stackrel{q=1}{\longrightarrow}\ P_{\Box}(A)^r, \nn\\
P_{[1^r]}(A) \ \stackrel{t=1}{\longrightarrow}\ P_{\Box}(A)^r
\ee
in the (anti)symmetric representations to the $\a$ case,
but, unfortunately, (\ref{sumain}) does not simplify enough.
Still, at the point $qt=1$ the number of terms in (\ref{sumain})
drops from $189$ to just $41$; this is still more than
$27=3^3$, which a cube of the fundamental superpolynomial
has at this point.

Nevertheless, at least in principle, one could continue searching
for a superpolynomial which satisfies an additional
property like

{\bf (G?)} At a subspace $f_R(q,t)=1$
(for example, $q^{r-1}t^{s-1}=1$ for the hook diagram)
\be
P_R(A|q,t) \ \stackrel{f_R(q,t)=1}{\longrightarrow}\
\prod_{(i,j)\in R} P_{_\Box}(Aq^{i-1}t^{1-j}\big|\,q,t) \ \ {\Large ?}
\label{facto}
\ee
\noindent
despite we did not find a way to satisfy it even for the
$\a$ representation.

\bigskip

However, it looks far more probable that the reality is
more interesting.
The key point is the apparent contradiction between
(\ref{facto}) and (\ref{ant2}).
If one expands the r.h.s. of (\ref{facto}) in powers of $\hbar$
and $\bar\hbar$, the first term would be
\be
\prod_{(i,j)\in R} P_{_\Box}^{{\cal K}}(Aq^{i-1}t^{1-j}\big|\,q,t)
=
\left(P_{_\Box}^{{\cal K}}(A|e^\hbar,e^{\bar\hbar})\right)^{|R|}
+ (\hbar\nu_{R'} - \bar\hbar\nu_{R})\sigma_1^{|R|-1}(A)
{P_{\Box}^{\cal K}}^\prime(A)
+ O(\hbar^2,\bar\hbar^2,\hbar\bar\hbar)
\ee
The main difference with (\ref{ant2}) is that the logarithmic $A$-derivative
$P_{\Box}'(A)/P_{_\Box}(A)=\sigma_1'(A)/\sigma_1(A) + O(\hbar,\bar\hbar)$
appears instead of $\sigma_2(A)/\sigma_1(A)^2$.
In particular, for the figure eight knot ${\cal K}=4_1$
one would get
\be
P_{[1]}^{4_1}(A)P_{[1]}^{4_1}(qA)P_{[1]}^{4_1}(At^{-1}) =
1+(\hbar-\bar\hbar)\{A^2\}\big(3\sigma_1^2+\sigma_1^2\sigma_1'(A)\big)
+ O(\hbar^2,\bar\hbar^2,\hbar\bar\hbar)= \nn \\
1+(\hbar-\bar\hbar)\{A^2\}\Big(5+10\{A\}^2+5\{A\}^4\Big)
+\ldots
\label{antiwr}
\ee
instead of $\big(5+12\{A\}^2+7\{A\}^4\big)$ in (\ref{anti}),
simply because $\sigma_1\sigma_1'(A) = 2\{A^2\}\big(1+\{A\}^2\big)$,
while $\sigma_2\{A\} = 2\{A^2\}\big(1+2\{A\}^2\big)$.
Note in passing that the difference is in higher order terms in $\{A\}^2$,
i.e. in higher order terms in $\epsilon^2$ whenever $A=q^N$.

From \cite{MMpols,Sle} we actually know what this substitution
$\sigma_1' \longrightarrow \sigma_2$ means:
in the world of knot polynomials the naive shift operators
$q^{d/d\log A}$ and $t^{-d/d\log A}$ are substituted by a somewhat more
complicated action of the generic cut-and-join operators \cite{MMN}
on {\it extended} knot polynomials \cite{MMM}
(which actually depend on time-variables and are expressed through $A$
only on the topological locus).
For $\bar\hbar=\hbar$, in the $\hbar$-linear approximation the action of this
$W$-evolution operator shifts $P^{\cal K}_R$ exactly by
$\hbar(\nu_{R'}-\nu_R)\sigma_{|R|-2}\sigma_2$
instead of $\hbar(\nu_{R'}-\nu_R)\sigma_{|R|-1}\sigma_1'$.
The combination $\varkappa_R = \nu_{R'}-\nu_R = \varphi_R([2])$
is the eigenvalue of the simplest cut-and-join operator $\hat W([2])$
on the $SL(\infty)$ character $\chi_R$ \cite{MMN}.
For $\bar\hbar\neq \hbar$ one needs a MacDonald generalization of these
operators, satisfying
\be
\hat{{\cal W}}(\Delta) M_R = \nu_{R'}(\Delta) M_R, \nn \\
{\hat{\bar {\cal W}}}(\Delta) M_R = \nu_{R}(\Delta) M_R
\ee
with $M_R$ being the MacDonald polynomials,
so that for $\Delta=[2]$ these MacDonald characters $\nu_R([2])$
coincide with the ordinary $\nu_R$ and in general provide a
proper decomposition of the symmetric group characters $\varphi_R(\Delta)$.
It is clear that such MacDonald version of the cut-and-join operators
will act on $P_R^{\cal K}$ by a shift
$(\hbar\nu_{R'}-\bar\hbar\nu_R)\sigma_2(A)/\sigma_1(A)$,
which is non-trivial already for $R=\Box$,
exactly as in (\ref{ant2}).

Thus an appropriate version of the factorization property (\ref{facto}),
which would provide a proper extension of \cite{Ant1} from the
(anti)symmetric representation, should contain an action of
(a MacDonald or refined version of) the $W$-evolution operator of \cite{Sle}
\be
\hspace{-1.5cm}{\bf (G)}:\hspace{4.5cm}
P_R(A|q,t) \ \stackrel{f_R(q,t)=1}{\longrightarrow}\
\left.\exp\left(\hat{{\cal W}}_R(q,t)\right)  {\cal P}_{_\Box}^{\otimes |R|}\{p\}
\right|_{p_k = \{A^k\}/\{t^k\}}
\ee
In the particular case of (anti)symmetric representations
\be
\hat{{\cal W}}_{[r]}(q=1,t) = 0,\nn \\
\hat{{\cal W}}_{[1^r]}(q,t=1) = 0
\ee
and this would explain why in this case the factorization in \cite{Ant1}
is so simple.
For other representation, however, the reduction of $\hat{\cal W}$ is not
so simple and remains to be worked out on the lines of \cite{Sle} and \cite{MMN}.

This problem has, of course, a lot in common with understanding
puzzles of the Ooguri-Vafa expansion \cite{OV} and its refined ($t$-deformed) version.

\section{Conclusion}

To conclude, in this paper we provide an explicit answer (\ref{main}) for
$H_{[21]}^{4_1}$, the  first non-trivial
colored HOMFLY polynomial for a non-torus knot in a non-(anti)symmetric
representation, as well as $H_{[21]}$ for some other 3-strand
knots, not obligatory twist ones (see the Appendix).

We also discuss a superpolynomial
$P_{[21]}^{4_1}$ and various ambiguities encountered in its construction.
In particular, we discuss a self-dual point where the superpolynomial drastically simplifies though not enough
to provide a factorization formula like \cite{Ant1}.
This once again emphasizes  the difficulties
still present in the superpolynomial theory.
Clearly, a universal object (superpolynomial or a variety of
superpolynomials) does exist, but its exact meaning and nature
still escapes us: this is what makes the subject
so interesting and appealing.

\section*{Acknowledgements}

Our work is partly supported by Ministry of Education and Science of
the Russian Federation under contract 8498 (A.A., A.Mir. and A.Mor) and 8606 (And.Mor.),
by CNPq 400635/2012-7, the Brazil National Counsel of Scientific and
Technological Development (A.Mor.), by the program  of UFRN-MCTI (Brazil) (A.Mir.), by NSh-3349.2012.2,
by RFBR grants 10-01-00536 (A.A., A.Mir., A.Mor.) and 10-02-01315 (And.Mor.),
by joint grants 11-02-90453-Ukr,
11-01-92612-Royal Society, 12-02-92108-Yaf-a  and by junior grant 12-02-31078 (A.A. and And.Mor.).

\newpage

\section*{Appendix. Cabling procedure}

We briefly explain here how to make the exact evaluation of the
colored HOMFLY polynomial using the cabling procedure within the
approach developed in \cite{MMM,AMMM} and list a few first examples
of the 3-strand knots in representation $\a$ evaluated with this
procedure for an illustrative purpose. This Appendix is an excerpt
from paper \cite{Ancab}, further details and the list of all other
3-strand knots with up to 8 crossings from the Rolfsen tables can be
found there.

\subsection*{Description of approach}

The colored HOMFLY polynomial of the knot $K$ can be evaluated as a linear combination
of the fundamental HOMFLY polynomials of several (more complicated)
knots and links. The main idea is that in order to evaluate the colored HOMFLY polynomial in the representation $Q$,
one has to look at the knot/link with each strand being substituted with a bunch of $p=|Q|$ strands ($p$-cabling of $K$). Then,
\be
{\cal H}_{[1]^{\otimes p}}(K)={\cal H}_{[1]}(K^p)=\sum_Q{\cal H}_{Q}(K^p),\ \ \ \ \ \ \ \ \ \f^{\otimes p}=\oplus Q
\ee
These
$p$ strands can additionally cross, and linear combinations of the crossings correspond to projectors onto
irreducible representations $Q$. There are at least two approaches to construct these projectors \cite{Ancab}.

The method that we use here exploits the idea that the form of this projector should not depend on the knot we are looking at
but only on the representation we are studying. Thus, to find the form of the projector one can look at the
simplest of knots, the unknot. One can represent the unknot in representation $Q$ in two different ways. On one hand,
the corresponding HOMFLY is equal to the $\SS_Q$. On the other hand, it can be represented as a sum of several knots
and links in fundamental representations with $|Q|$ intertwining strands with some coefficient. From these two representations
one can construct the projectors.

\subsubsection*{Example: $Q=\d$ and $Q=\e$}

For instance, in the case of $2$-cable unknot one has two possibilities: the two strands can go
without crossings (${\cal H}^{(0)}=\SS_{[2]}+\SS_{[11]}$) or can cross once (${\cal H}^{(1)}=\SS_{[2]} q-\SS_{[11]} q^{-1}$).
Solving the system

\begin{eqnarray}
\begin{array}{c}
\SS_{[2]}=p_{[2]}^{0}{\cal H}^{(0)}+p_{[2]}^{1}{\cal H}^{(1)}
\\
\SS_{[11]}=p_{[11]}^{0}{\cal H}^{(0)}+p_{[11]}^{1}{\cal H}^{(1)}
\end{array}
\end{eqnarray}
one finds the projectors
\begin{eqnarray}
\begin{array}{llll}
p_{[2]}^0= \frac{1}{q(q+q^{-1})}& & &p_{[2]}^1= \frac{1}{(q+q^{-1})}
\\
\\
p_{[11]}^0= \frac{q}{(q+q^{-1})}& & &p_{[11]}^1= -\frac{1}{(q+q^{-1})}
\end{array}
\end{eqnarray}
One can reformulate the calculation in terms of the braid representation of the knot and
the $\fR$-matrix realizing the generators of the braid group \cite{RT,MMM},
the projectors in the terms of the ${\fR}$-matrix acting in $\f\otimes \f$ being
\cite{MMpols}
\begin{eqnarray}
\begin{array}{l}
P_{[2]}= \frac{1}{q(q+q^{-1})}+\frac{1}{(q+q^{-1})}{\fR}
\\
\\
P_{[11]}= \frac{q}{(q+q^{-1})}-\frac{1}{(q+q^{-1})}{\fR}
\end{array}
\end{eqnarray}
so that $P_2^2=P_2$, $P_2P_{11}=0$ and $P_{11}^2=P_{11}$ due to the skein relation
for the fundamental $\fR$-matrix: $\fR^2 = 1 + (q-q^{-1})\fR$.

\subsubsection*{Example: $Q=\b$, $Q=\a$ and $Q=\c$}

The next case is 3-cabling and 3 irreducible representations $\b$, $\a$ and $\c$.
One again expresses the unknot in the terms of $3$ strands without crossings
 $\H^{(00)}$ (three unknots), with one crossing $\H^{(10)}$ (two unknots) and with one crossing between
two strands and another one between two other crossings: $\H^{(11)}$ (one unknot).
These are manifestly given by expressions
\begin{eqnarray}
\H^{(00)}=\SS_{[3]}+2\SS_{[21]}+\SS_{[111]}\nn
\\
\H^{(10)}=q\SS_{[3]}+\left(q-q^{-1}\right)\SS_{[21]}-q^{-1}\SS_{[111]}\nn
\\
\H^{(11)}=q^2\SS_{[3]}-\SS_{[21]}+q^{-2}\SS_{[111]}
\end{eqnarray}
i.e. the three projectors are (the $\R$-matrices are here colored, i.e. taken in non-fundamental representations)
\begin{eqnarray}\label{prans}
\begin{array}{l}
P_{[3]}=
{1\over q^3[2][3]}\Big(1 + 2q \R_1+2q^2 \R_1\R_2 + q^3\R_1\R_2\R_1\Big)
\\
\\
P_{[21]}=\frac{1}{q^2+1+q^{-2}}\Big(1+(q-q^{-1})\R_1-\R_1\R_2\Big)
\\
\\
P_{[111]}=
-{1\over [2][3]}\Big(q^3 + 2q^2 \R_1+2q \R_1\R_2 + \R_1\R_2\R_1\Big)
\end{array}
\end{eqnarray}
Here we are most interested in the representation $\a$, the corresponding projector $P_{[21]}$ can be represented
as
\be\label{proj}
P_{[21]}=\frac{1}{q^2+1+q^{-2}}\left(\R_1^2-\R_1\R_2\right)
\ee
because
\be
\H^{(20)}=q^2\SS_{[3]}+(q^2+q^{-2})\SS_{[21]}+q^{-2}\SS_{[111]}
\ee

\subsection*{Representation $\a$}

In order to calculate the HOMFLY polynomial of 3-strand knots, one has to substitute in the 3-strand braid
describing the knot all the strands by triple strands, accordingly increasing the number of intersections.
Thus, one obtains a 9-strand braid which is dealt with along the line of \cite{AMMM}. In other words,
the colored $\R$-matrices in the original 3-strand braid should be substituted by products of
fundamental $\fR$-matrices
in the 9-strand braid in accordance with the rule
\begin{eqnarray}
\label{repcr}
\begin{array}{l}
\R_1\rightarrow {\R}_1=\fR_3\fR_2\fR_1 \fR_4\fR_3\fR_2 \fR_5\fR_4\fR_3
\\
\R_2\rightarrow {\R}_2=\fR_6\fR_5\fR_4 \fR_7\fR_6\fR_5 \fR_8\fR_7\fR_6
\end{array}
\end{eqnarray}
As usual $\R_i$ denotes the ${\R}$-matrix acting on the crossing of $i$-th and $i+1$-th strands in the braid,
i.e. $\R_{1}$ and $\R_2$ act on $V\otimes V\otimes I$ and $I\otimes V\otimes V$ respectively, while $\fR_i$ acts
on $\underbrace{I\otimes\ldots\otimes I}_{i-1}\otimes V\otimes V\otimes I\ldots\otimes I$.

Thus, to evaluate the $3$-strand knots for representation $\a$ one has
to evaluate the $9$-strand knots in
the fundamental representation and project the answer using (\ref{proj}) \cite{Ancab}. The results of calculations are present
below. In fact, at the 9-strand level one can also similarly obtain the answers
for representations $\b$, $\a$ and $\c$. We, indeed, made these calculations and checked that the answers of ref.\cite{IMMM3}
for $\b$ and $\c$ are reproduced.
This is a non-trivial check, because the cabling calculation
is based on the very reliable conjecture of \cite{AMMM}
about the mixing  (Racah) matrices in the fundamental
representation, while the calculation in \cite{IMMM3}
is based on a far less reliable "eigenvalue" conjecture about
the mixing matrices in non-trivial representations.
Thus, this coincidence not only checks our calculations,
but provides a strong support to the reasoning of \cite{IMMM3}.

\subsection*{A list of $\H_{[21]}$ for simplest $3$-strand knots \cite{Ancab}}

\bigskip

\section*{$\boxed{5_2}$}
{\footnotesize
\be
A^{-6}&
-1\nn\\
A^{-4}&
q^{6}+q^{2}+q^{-2}+q^{-6}-1\nn\\
A^{-2}&
q^{6}+q^{-6}-2\nn\\
1&
-q^{14}+q^{12}-3\,q^{10}+3\,q^{8}-5\,q^{6}+7\,q^{4}-8\,q
^{2}-8\,q^{-2}+7\,q^{-4}-5\,q^{-6}+3\,q^{-8}-\nn\\&-3\,q^{-10}+q^
{-12}-q^{-14}+7\nn\\
A^{2}&
q^{14}-2\,q^{12}+4\,q^{10}-6\,q^{8}+9\,q^{6}-12\,q^{4}+13
\,q^{2}+13\,q^{-2}-12\,q^{-4}+9\,q^{-6}-\nn\\&-6\,q^{-8}+4\,q^{-
10}-2\,q^{-12}+q^{-14}-14\nn\\
A^{4}&
q^{12}-2\,q^{10}+5\,q^{8}-7\,q^{6}+9\,q^{4}-11\,q^{2}-11\,
q^{-2}+9\,q^{-4}-7\,q^{-6}+5\,q^{-8}-\nn\\&-2\,q^{-10}+q^{-12}+13\nn\\
A^{6}&
q^{10}-2\,q^{8}+3\,q^{6}-4\,q^{4}+5\,q^{2}+5\,q^{-2}-4\,{q
}^{-4}+3\,q^{-6}-2\,q^{-8}+q^{-10}-5\nn\ee
}

\section*{$\boxed{6_{2}}$}
{\footnotesize
\be
A^{-6}&
q^{10}+2\,q^{6}-q^{4}+2\,q^{2}+2\,q^{-2}-q^{-4}+2\,q^{-6
}+q^{-10}\nn\\
A^{-4}&
-q^{16}-3\,q^{12}+2\,q^{10}-5\,q^{8}+3\,q^{6}-8\,q^{4}+4\,
q^{2}+4\,q^{-2}-8\,q^{-4}+3\,q^{-6}-\nn\\&-5\,q^{-8}+2\,q^{-10}-3
\,q^{-12}-q^{-16}-8\nn\\
A^{-2}&
q^{20}-q^{18}+5\,q^{16}-7\,q^{14}+13\,q^{12}-14\,q^{10}+24
\,q^{8}-22\,q^{6}+29\,q^{4}-26\,q^{2}-26\,q^{-2}+\nn\\&+29\,q^{-4
}-22\,q^{-6}+24\,q^{-8}-14\,q^{-10}+13\,q^{-12}-\nn\\&-7\,q^{-14}+5
\,q^{-16}-q^{-18}+q^{-20}+32\nn\\
1&
-2\,q^{20}+3\,q^{18}-8\,q^{16}+12\,q^{14}-22\,q^{12}+24\,q
^{10}-33\,q^{8}+35\,q^{6}-42\,q^{4}+39\,q^{2}+\nn\\&+39\,q^{-2}-42
\,q^{-4}+35\,q^{-6}-33\,q^{-8}+24\,q^{-10}-\nn\\&-22\,q^{-12}+12\,{
q}^{-14}-8\,q^{-16}+3\,q^{-18}-2\,q^{-20}-44\nn\\
A^{2}&
q^{20}-2\,q^{18}+5\,q^{16}-6\,q^{14}+12\,q^{12}-14\,q^{10}
+17\,q^{8}-16\,q^{6}+21\,q^{4}-18\,q^{2}-18\,q^{-2}+\nn\\&+21\,q^
{-4}-16\,q^{-6}+17\,q^{-8}-14\,q^{-10}+12\,q^{-12}-6\,q^{-14
}+5\,q^{-16}-2\,q^{-18}+q^{-20}+18\nn\\
A^{4}&
-q^{16}+q^{14}-q^{12}-q^{8}-2\,q^{6}+5\,q^{4}-6\,q^{2}-6
\,q^{-2}+5\,q^{-4}-2\,q^{-6}-q^{-8}-\nn\\&-q^{-12}+q^{-14}-q^{-
16}+4\nn\\
A^{6}&
q^{10}-2\,q^{8}+3\,q^{6}-4\,q^{4}+5\,q^{2}+5\,q^{-2}-4\,{q
}^{-4}+3\,q^{-6}-2\,q^{-8}+q^{-10}-5\nn\ee
}
\section*{$\boxed{6_{3}}$}
{\footnotesize
\be
A^{-6}&
-q^{10}+2\,q^{8}-3\,q^{6}+4\,q^{4}-5\,q^{2}-5\,q^{-2}+4\,{
q}^{-4}-3\,q^{-6}+2\,q^{-8}-q^{-10}+5\nn\\
A^{-4}&
q^{16}-q^{14}+2\,q^{12}-2\,q^{10}+4\,q^{8}-4\,q^{6}+6\,q
^{4}-5\,q^{2}-5\,q^{-2}+6\,q^{-4}-4\,q^{-6}+\nn\\&+4\,q^{-8}-2\,q
^{-10}+2\,q^{-12}-q^{-14}+q^{-16}+7\nn\\
A^{-2}&
-q^{20}+2\,q^{18}-6\,q^{16}+9\,q^{14}-17\,q^{12}+23\,q^{10
}-36\,q^{8}+41\,q^{6}-55\,q^{4}+\nn\\&+56\,q^{2}+56\,q^{-2}-55\,q
^{-4}+41\,q^{-6}-36\,q^{-8}+23\,q^{-10}-\nn\\&-17\,q^{-12}+9\,q^{-
14}-6\,q^{-16}+2\,q^{-18}-q^{-20}-62\nn\\
1&
2\,q^{20}-4\,q^{18}+10\,q^{16}-16\,q^{14}+31\,q^{12}-40\,q
^{10}+60\,q^{8}-71\,q^{6}+90\,q^{4}-\nn\\&-92\,q^{2}-92\,q^{-2}+90
\,q^{-4}-71\,q^{-6}+60\,q^{-8}-40\,q^{-10}+31\,q^{-12}-\nn\\&-16\,{
q}^{-14}+10\,q^{-16}-4\,q^{-18}+2\,q^{-20}+105\nn\\
A^{2}&
-q^{20}+2\,q^{18}-6\,q^{16}+9\,q^{14}-17\,q^{12}+23\,q^{10
}-36\,q^{8}+41\,q^{6}-55\,q^{4}+56\,q^{2}+\nn\\&+56\,q^{-2}-55\,q
^{-4}+41\,q^{-6}-36\,q^{-8}+23\,q^{-10}-17\,q^{-12}+9\,q^{-
14}-6\,q^{-16}+\nn\\&+2\,q^{-18}-q^{-20}-62\nn\\
A^{4}&
q^{16}-q^{14}+2\,q^{12}-2\,q^{10}+4\,q^{8}-4\,q^{6}+6\,q
^{4}-5\,q^{2}-5\,q^{-2}+6\,q^{-4}-4\,q^{-6}+4\,q^{-8}-\nn\\&-2\,q
^{-10}+2\,q^{-12}-q^{-14}+q^{-16}+7\nn\\
A^{6}&
-q^{10}+2\,q^{8}-3\,q^{6}+4\,q^{4}-5\,q^{2}-5\,q^{-2}+4\,{
q}^{-4}-3\,q^{-6}+2\,q^{-8}-q^{-10}+5\nn\ee
}
\section*{$\boxed{7_{3}}$}

\vspace{-1.3cm}
{\footnotesize
\be
A^{-6}&
-q^{10}-2\,q^{6}+q^{4}-2\,q^{2}-2\,q^{-2}+q^{-4}-2\,q^{-
6}-q^{-10}\nn\\
A^{-4}&
q^{16}+3\,q^{12}-2\,q^{10}+5\,q^{8}-3\,q^{6}+8\,q^{4}-4\,{
q}^{2}-4\,q^{-2}+8\,q^{-4}-3\,q^{-6}+5\,q^{-8}-\nn\\&-2\,q^{-10}+3
\,q^{-12}+q^{-16}+8\nn\\
A^{-2}&
q^{14}-2\,q^{12}+3\,q^{10}-8\,q^{8}+9\,q^{6}-13\,q^{4}+13
\,q^{2}+13\,q^{-2}-13\,q^{-4}+9\,q^{-6}-8\,q^{-8}+\nn\\&+3\,q^{-
10}-2\,q^{-12}+q^{-14}-18\nn\\
1&
-q^{24}+q^{22}-4\,q^{20}+5\,q^{18}-9\,q^{16}+10\,q^{14}-14
\,q^{12}+13\,q^{10}-14\,q^{8}+11\,q^{6}-10\,q^{4}+\nn\\&+7\,q^{2}
+7\,q^{-2}-10\,q^{-4}+11\,q^{-6}-14\,q^{-8}+13\,q^{-10}-\nn\\&-14\,
q^{-12}+10\,q^{-14}-9\,q^{-16}+5\,q^{-18}-4\,q^{-20}+q^{-
22}-q^{-24}-6\nn\\
A^{2}&
q^{24}-2\,q^{22}+5\,q^{20}-9\,q^{18}+16\,q^{16}-22\,q^{14}
+28\,q^{12}-33\,q^{10}+40\,q^{8}-41\,q^{6}+39\,q^{4}-\nn\\&-41\,q
^{2}-41\,q^{-2}+39\,q^{-4}-41\,q^{-6}+40\,q^{-8}-\nn\\&-33\,q^{-10}
+28\,q^{-12}-22\,q^{-14}+16\,q^{-16}-9\,q^{-18}+5\,q^{-20}-2
\,q^{-22}+q^{-24}+44\nn\\
A^{4}&
q^{22}-2\,q^{20}+6\,q^{18}-11\,q^{16}+17\,q^{14}-23\,q^{12
}+31\,q^{10}-35\,q^{8}+38\,q^{6}-41\,q^{4}+42\,q^{2}+\nn\\&+42\,q
^{-2}-41\,q^{-4}+38\,q^{-6}-35\,q^{-8}+31\,q^{-10}-23\,q^{-
12}+\nn\\&+17\,q^{-14}-11\,q^{-16}+6\,q^{-18}-2\,q^{-20}+q^{-22}-40\nn\\
A^{6}&
q^{20}-2\,q^{18}+3\,q^{16}-6\,q^{14}+10\,q^{12}-11\,q^{10}
+12\,q^{8}-15\,q^{6}+16\,q^{4}-15\,q^{2}-15\,q^{-2}+\nn\\&+16\,q^
{-4}-15\,q^{-6}+12\,q^{-8}-11\,q^{-10}+10\,q^{-12}-6\,q^{-14
}+3\,q^{-16}-2\,q^{-18}+q^{-20}+15\nn\ee
}
\section*{$\boxed{7_{5}}$}

\vspace{-1.3cm}
{\footnotesize
\be
A^{-6}&
q^{20}-2\,q^{18}+5\,q^{16}-9\,q^{14}+15\,q^{12}-20\,q^{10}
+27\,q^{8}-32\,q^{6}+38\,q^{4}-40\,q^{2}-40\,q^{-2}+\nn\\&+38\,q^
{-4}-32\,q^{-6}+27\,q^{-8}-20\,q^{-10}+15\,q^{-12}-9\,q^{-14
}+5\,q^{-16}-2\,q^{-18}+q^{-20}+42\nn\\
A^{-4}&
q^{22}-3\,q^{20}+9\,q^{18}-19\,q^{16}+33\,q^{14}-49\,q^{12
}+69\,q^{10}-90\,q^{8}+107\,q^{6}-121\,q^{4}+\nn\\&+130\,q^{2}+130
\,q^{-2}-121\,q^{-4}+107\,q^{-6}-90\,q^{-8}+69\,q^{-10}-49\,
q^{-12}+\nn\\&+33\,q^{-14}-19\,q^{-16}+9\,q^{-18}-3\,q^{-20}+q^{-
22}-134\nn\\
A^{-2}&
q^{24}-3\,q^{22}+8\,q^{20}-18\,q^{18}+30\,q^{16}-48\,q^{14
}+69\,q^{12}-92\,q^{10}+111\,q^{8}-133\,q^{6}+\nn\\&+146\,q^{4}-156
\,q^{2}-156\,q^{-2}+146\,q^{-4}-133\,q^{-6}+111\,q^{-8}-92\,
q^{-10}+69\,q^{-12}-\nn\\&-48\,q^{-14}+30\,q^{-16}-18\,q^{-18}+8\,{
q}^{-20}-3\,q^{-22}+q^{-24}+158\nn\\
1&
-q^{24}+2\,q^{22}-6\,q^{20}+10\,q^{18}-17\,q^{16}+24\,q^{
14}-32\,q^{12}+41\,q^{10}-47\,q^{8}+51\,q^{6}-\nn\\&-55\,q^{4}+58\,
q^{2}+58\,q^{-2}-55\,q^{-4}+51\,q^{-6}-47\,q^{-8}+41\,q^{-
10}-32\,q^{-12}+\nn\\&+24\,q^{-14}-17\,q^{-16}+10\,q^{-18}-6\,q^{-
20}+2\,q^{-22}-q^{-24}-56\nn\\
A^{2}&
q^{18}+2\,q^{14}-4\,q^{12}+9\,q^{10}-12\,q^{8}+17\,q^{6}-
24\,q^{4}+27\,q^{2}+27\,q^{-2}-24\,q^{-4}+\nn\\&+17\,q^{-6}-12\,q
^{-8}+9\,q^{-10}-4\,q^{-12}+2\,q^{-14}+q^{-18}-26\nn\\
A^{4}&
q^{16}-2\,q^{14}+3\,q^{12}-6\,q^{10}+9\,q^{8}-11\,q^{6}+12
\,q^{4}-14\,q^{2}-14\,q^{-2}+12\,q^{-4}-\nn\\&-11\,q^{-6}+9\,q^{-
8}-6\,q^{-10}+3\,q^{-12}-2\,q^{-14}+q^{-16}+16\nn\\
A^{6}&
-q^{10}+2\,q^{8}-3\,q^{6}+4\,q^{4}-5\,q^{2}-5\,q^{-2}+4\,{
q}^{-4}-3\,q^{-6}+2\,q^{-8}-q^{-10}+5\nn\ee
}
\section*{$\boxed{8_{2}}$}

\vspace{-1.3cm}
{\footnotesize
\be
A^{-6}&
q^{20}+2\,q^{16}-q^{14}+4\,q^{12}-q^{10}+4\,q^{8}-q^{6}+
4\,q^{4}-q^{2}-q^{-2}+4\,q^{-4}-q^{-6}+4\,q^{-8}-\nn\\&-q^{-10}
+4\,q^{-12}-q^{-14}+2\,q^{-16}+q^{-20}+5\nn\\
A^{-4}&
-q^{26}-3\,q^{22}+2\,q^{20}-7\,q^{18}+4\,q^{16}-12\,q^{14}
+7\,q^{12}-15\,q^{10}+8\,q^{8}-18\,q^{6}+8\,q^{4}-\nn\\&-18\,q^{2
}-18\,q^{-2}+8\,q^{-4}-18\,q^{-6}+8\,q^{-8}-15\,q^{-10}+7\,{
q}^{-12}-12\,q^{-14}+\nn\\&+4\,q^{-16}-7\,q^{-18}+2\,q^{-20}-3\,q^{
-22}-q^{-26}+9\nn\\
A^{-2}&
q^{30}-q^{28}+5\,q^{26}-7\,q^{24}+15\,q^{22}-18\,q^{20}+32
\,q^{18}-32\,q^{16}+46\,q^{14}-43\,q^{12}+\+57\,q^{10}-\nn\\&-48\,q
^{8}+61\,q^{6}-52\,q^{4}+64\,q^{2}+64\,q^{-2}-52\,q^{-4}+61
\,q^{-6}-48\,q^{-8}+57\,q^{-10}-\nn\\&-43\,q^{-12}+46\,q^{-14}-32\,
q^{-16}+32\,q^{-18}-18\,q^{-20}+15\,q^{-22}-7\,q^{-24}+5\,{q
}^{-26}-q^{-28}+q^{-30}-52\nn\\
1&
-2\,q^{30}+3\,q^{28}-8\,q^{26}+13\,q^{24}-25\,q^{22}+30\,q
^{20}-45\,q^{18}+50\,q^{16}-62\,q^{14}+58\,q^{12}-\nn\\&-69\,q^{10}
+64\,q^{8}-71\,q^{6}+62\,q^{4}-71\,q^{2}-71\,q^{-2}+62\,q^
{-4}-71\,q^{-6}+64\,q^{-8}-69\,q^{-10}+\nn\\&+58\,q^{-12}-62\,q^{-
14}+50\,q^{-16}-45\,q^{-18}+30\,q^{-20}-25\,q^{-22}+13\,q^{-
24}-8\,q^{-26}+3\,q^{-28}-2\,q^{-30}+65\nn\\
A^{2}&
q^{30}-2\,q^{28}+5\,q^{26}-7\,q^{24}+14\,q^{22}-17\,q^{20}
+22\,q^{18}-22\,q^{16}+26\,q^{14}-20\,q^{12}+\nn\\&+19\,q^{10}-12\,
q^{8}+12\,q^{6}-7\,q^{4}+8\,q^{2}+8\,q^{-2}-7\,q^{-4}+12\,
q^{-6}-12\,q^{-8}+19\,q^{-10}-20\,q^{-12}+\nn\\&+26\,q^{-14}-22\,{q
}^{-16}+22\,q^{-18}-17\,q^{-20}+14\,q^{-22}-7\,q^{-24}+5\,q^
{-26}-2\,q^{-28}+q^{-30}-4\nn\\
A^{4}&
-q^{26}+q^{24}-q^{22}+q^{20}-q^{18}-5\,q^{16}+9\,q^{14}-
13\,q^{12}+20\,q^{10}-28\,q^{8}+29\,q^{6}-\nn\\&-31\,q^{4}+33\,q^
{2}+33\,q^{-2}-31\,q^{-4}+29\,q^{-6}-28\,q^{-8}+20\,q^{-10}-
13\,q^{-12}+9\,q^{-14}-5\,q^{-16}-\nn\\&-q^{-18}+q^{-20}-q^{-22}+
q^{-24}-q^{-26}-35\nn\\
A^{6}&
q^{20}-2\,q^{18}+3\,q^{16}-6\,q^{14}+10\,q^{12}-11\,q^{10}
+12\,q^{8}-15\,q^{6}+16\,q^{4}-15\,q^{2}-15\,q^{-2}+\nn\\&+16\,q^
{-4}-15\,q^{-6}+12\,q^{-8}-11\,q^{-10}+10\,q^{-12}-6\,q^{-14
}+3\,q^{-16}-2\,q^{-18}+q^{-20}+15\nn\ee
}
\section*{$\boxed{8_{5}}$}
{\footnotesize
\be
A^{-6}&
q^{20}-2\,q^{18}+4\,q^{16}-6\,q^{14}+9\,q^{12}-8\,q^{10}+7
\,q^{8}-5\,q^{6}+5\,q^{4}-q^{2}-q^{-2}+5\,q^{-4}-\nn\\&-5\,q^{-
6}+7\,q^{-8}-8\,q^{-10}+9\,q^{-12}-6\,q^{-14}+4\,q^{-16}-2\,
q^{-18}+q^{-20}\nn\\
A^{-4}&
-q^{26}+q^{24}-2\,q^{22}-10\,q^{16}+14\,q^{14}-26\,q^{12}+
30\,q^{10}-46\,q^{8}+42\,q^{6}-49\,q^{4}+44\,q^{2}+\nn\\&+44\,q^{
-2}-49\,q^{-4}+42\,q^{-6}-46\,q^{-8}+30\,q^{-10}-26\,q^{-12}
+14\,q^{-14}-\nn\\&-10\,q^{-16}-2\,q^{-22}+q^{-24}-q^{-26}-54\nn\\
A^{-2}&
q^{30}-2\,q^{28}+6\,q^{26}-7\,q^{24}+16\,q^{22}-14\,q^{20}
+22\,q^{18}-11\,q^{16}+18\,q^{14}+8\,q^{12}+\nn\\&+30\,q^{8}-16\,{q
}^{6}+46\,q^{4}-23\,q^{2}-23\,q^{-2}+46\,q^{-4}-16\,q^{-6}+
30\,q^{-8}+\nn\\&+8\,q^{-12}+18\,q^{-14}-11\,q^{-16}+22\,q^{-18}-14
\,q^{-20}+16\,q^{-22}-7\,q^{-24}+6\,q^{-26}-2\,q^{-28}+q^{
-30}+50\nn\\
1&
-2\,q^{30}+3\,q^{28}-11\,q^{26}+16\,q^{24}-35\,q^{22}+40\,{q
}^{20}-68\,q^{18}+64\,q^{16}-97\,q^{14}+-\nn\\&+77\,q^{12}-115\,q^{
10}+83\,q^{8}-130\,q^{6}+84\,q^{4}-136\,q^{2}-136\,q^{-2}+84
\,q^{-4}-\nn\\&-130\,q^{-6}+83\,q^{-8}-115\,q^{-10}+77\,q^{-12}-97
\,q^{-14}+64\,q^{-16}-68\,q^{-18}+\nn\\&+40\,q^{-20}-35\,q^{-22}+16
\,q^{-24}-11\,q^{-26}+3\,q^{-28}-2\,q^{-30}+89\nn\\
A^{2}&
q^{30}-q^{28}+7\,q^{26}-10\,q^{24}+26\,q^{22}-31\,q^{20}+
61\,q^{18}-61\,q^{16}+102\,q^{14}-\nn\\&-91\,q^{12}+143\,q^{10}-114
\,q^{8}+173\,q^{6}-133\,q^{4}+194\,q^{2}+194\,q^{-2}-133\,{q
}^{-4}+\nn\\&+173\,q^{-6}-114\,q^{-8}+143\,q^{-10}-91\,q^{-12}+102\,{
q}^{-14}-61\,q^{-16}+61\,q^{-18}-\nn\\&-31\,q^{-20}+26\,q^{-22}-10\,{
q}^{-24}+7\,q^{-26}-q^{-28}+q^{-30}-136\nn\\
A^{4}&
-q^{26}-5\,q^{22}+3\,q^{20}-14\,q^{18}+10\,q^{16}-29\,q^{
14}+18\,q^{12}-46\,q^{10}+27\,q^{8}-62\,q^{6}+\nn\\&+31\,q^{4}-70\,
q^{2}-70\,q^{-2}+31\,q^{-4}-62\,q^{-6}+27\,q^{-8}-46\,q^{-
10}+18\,q^{-12}-\nn\\&-29\,q^{-14}+10\,q^{-16}-14\,q^{-18}+3\,q^{-
20}-5\,q^{-22}-q^{-26}+36\nn\\
A^{6}&
q^{20}+4\,q^{16}-2\,q^{14}+8\,q^{12}-4\,q^{10}+13\,q^{8}-6
\,q^{6}+16\,q^{4}-8\,q^{2}-8\,q^{-2}+16\,q^{-4}-\nn\\&-6\,q^{-6}+
13\,q^{-8}-4\,q^{-10}+8\,q^{-12}-2\,q^{-14}+4\,q^{-16}+q^{
-20}+20\nn\ee
}
\section*{$\boxed{8_{7}}$}
{\footnotesize
\be
A^{-6}&
-q^{20}+2\,q^{18}-5\,q^{16}+9\,q^{14}-15\,q^{12}+20\,q^{10
}-27\,q^{8}+32\,q^{6}-38\,q^{4}+40\,q^{2}+40\,q^{-2}-\nn\\&-38\,q
^{-4}+32\,q^{-6}-27\,q^{-8}+20\,q^{-10}-15\,q^{-12}+9\,q^{-
14}-5\,q^{-16}+2\,q^{-18}-q^{-20}-42\nn\\
A^{-4}&
q^{26}-q^{24}+3\,q^{22}-2\,q^{20}+3\,q^{18}+2\,q^{16}-5\,{
q}^{14}+18\,q^{12}-25\,q^{10}+45\,q^{8}-56\,q^{6}+\nn\\&+75\,q^{4}-
78\,q^{2}-78\,q^{-2}+75\,q^{-4}-56\,q^{-6}+45\,q^{-8}-25\,{q
}^{-10}+18\,q^{-12}-\nn\\&-5\,q^{-14}+2\,q^{-16}+3\,q^{-18}-2\,q^{-
20}+3\,q^{-22}-q^{-24}+q^{-26}+88\nn\\
A^{-2}&
-q^{30}+2\,q^{28}-7\,q^{26}+11\,q^{24}-22\,q^{22}+31\,q^{
20}-49\,q^{18}+55\,q^{16}-76\,q^{14}+\nn\\&+78\,q^{12}-90\,q^{10}+
77\,q^{8}-85\,q^{6}+66\,q^{4}-72\,q^{2}-72\,q^{-2}+66\,q^{
-4}-\nn\\&-85\,q^{-6}+77\,q^{-8}-90\,q^{-10}+78\,q^{-12}-76\,q^{-14
}+55\,q^{-16}-49\,q^{-18}+\nn\\&+31\,q^{-20}-22\,q^{-22}+11\,q^{-24
}-7\,q^{-26}+2\,q^{-28}-q^{-30}+56\nn\\
1&
2\,q^{30}-4\,q^{28}+11\,q^{26}-20\,q^{24}+40\,q^{22}-57\,q
^{20}+91\,q^{18}-118\,q^{16}+\nn\\&+159\,q^{14}-181\,q^{12}+220\,q^
{10}-233\,q^{8}+260\,q^{6}-255\,q^{4}+\nn\\&+273\,q^{2}+273\,q^{-2}
-255\,q^{-4}+260\,q^{-6}-233\,q^{-8}+220\,q^{-10}-\nn\\&-181\,q^{-
12}+159\,q^{-14}-118\,q^{-16}+91\,q^{-18}-57\,q^{-20}+40\,q^
{-22}-\nn\\&-20\,q^{-24}+11\,q^{-26}-4\,q^{-28}+2\,q^{-30}-264\nn\\
A^{2}&
-q^{30}+2\,q^{28}-6\,q^{26}+11\,q^{24}-23\,q^{22}+34\,q^{
20}-55\,q^{18}+75\,q^{16}-\nn\\&-104\,q^{14}+124\,q^{12}-155\,q^{10
}+171\,q^{8}-190\,q^{6}+198\,q^{4}-\nn\\&-210\,q^{2}-210\,q^{-2}+
198\,q^{-4}-190\,q^{-6}+171\,q^{-8}-155\,q^{-10}+\nn\\&+124\,q^{-12
}-104\,q^{-14}+75\,q^{-16}-55\,q^{-18}+34\,q^{-20}-\nn\\&-23\,q^{-
22}+11\,q^{-24}-6\,q^{-26}+2\,q^{-28}-q^{-30}+204\nn\\
A^{4}&
q^{26}-q^{24}+2\,q^{22}-4\,q^{20}+7\,q^{18}-6\,q^{16}+11\,
q^{14}-17\,q^{12}+19\,q^{10}-21\,q^{8}+29\,q^{6}-\nn\\&-30\,q^{4}
+32\,q^{2}+32\,q^{-2}-30\,q^{-4}+29\,q^{-6}-21\,q^{-8}+19\,{
q}^{-10}-17\,q^{-12}+\nn\\&+11\,q^{-14}-6\,q^{-16}+7\,q^{-18}-4\,q^
{-20}+2\,q^{-22}-q^{-24}+q^{-26}-32\nn\\
A^{6}&
-q^{20}+2\,q^{18}-3\,q^{16}+6\,q^{14}-10\,q^{12}+11\,q^{10
}-12\,q^{8}+15\,q^{6}-16\,q^{4}+15\,q^{2}+15\,q^{-2}-\nn\\&-16\,q
^{-4}+15\,q^{-6}-12\,q^{-8}+11\,q^{-10}-10\,q^{-12}+6\,q^{-
14}-3\,q^{-16}+2\,q^{-18}-q^{-20}-15\nn\ee
}
\section*{$\boxed{8_{9}}$}

\vspace{-1.3cm}
{\footnotesize
\be
A^{-6}&
q^{20}-2\,q^{18}+5\,q^{16}-9\,q^{14}+15\,q^{12}-20\,q^{10}
+27\,q^{8}-32\,q^{6}+38\,q^{4}-40\,q^{2}-40\,q^{-2}+\nn\\&+38\,q^
{-4}-32\,q^{-6}+27\,q^{-8}-20\,q^{-10}+15\,q^{-12}-9\,q^{-14
}+5\,q^{-16}-2\,q^{-18}+q^{-20}+42\nn\\
A^{-4}&
-q^{26}+q^{24}-3\,q^{22}+3\,q^{20}-5\,q^{18}+q^{16}-q^{
14}-5\,q^{12}+8\,q^{10}-18\,q^{8}+16\,q^{6}-21\,q^{4}+\nn\\&+20\,{q
}^{2}+20\,q^{-2}-21\,q^{-4}+16\,q^{-6}-18\,q^{-8}+8\,q^{-10}
-5\,q^{-12}-q^{-14}+q^{-16}-\nn\\&-5\,q^{-18}+3\,q^{-20}-3\,q^{-
22}+q^{-24}-q^{-26}-26\nn\\
A^{-2}&
q^{30}-2\,q^{28}+7\,q^{26}-12\,q^{24}+25\,q^{22}-36\,q^{20
}+58\,q^{18}-74\,q^{16}+\nn\\&+103\,q^{14}-121\,q^{12}+155\,q^{10}-
175\,q^{8}+208\,q^{6}-224\,q^{4}+\nn\\&+247\,q^{2}+247\,q^{-2}-224
\,q^{-4}+208\,q^{-6}-175\,q^{-8}+155\,q^{-10}-\nn\\&-121\,q^{-12}+
103\,q^{-14}-74\,q^{-16}+58\,q^{-18}-36\,q^{-20}+25\,q^{-22}
-\nn\\&-12\,q^{-24}+7\,q^{-26}-2\,q^{-28}+q^{-30}-242\nn\\
1&
-2\,q^{30}+4\,q^{28}-12\,q^{26}+22\,q^{24}-44\,q^{22}+64\,{q
}^{20}-103\,q^{18}+\nn\\&+136\,q^{16}-186\,q^{14}+225\,q^{12}-286\,{q
}^{10}+332\,q^{8}-389\,q^{6}+\nn\\&+414\,q^{4}-454\,q^{2}-454\,q^{-
2}+414\,q^{-4}-389\,q^{-6}+332\,q^{-8}-\nn\\&-286\,q^{-10}+225\,q^{
-12}-186\,q^{-14}+136\,q^{-16}-103\,q^{-18}+64\,q^{-20}-44\,{q
}^{-22}+\nn\\&+22\,q^{-24}-12\,q^{-26}+4\,q^{-28}-2\,q^{-30}+459\nn\\
A^{2}&
q^{30}-2\,q^{28}+7\,q^{26}-12\,q^{24}+25\,q^{22}-36\,q^{20
}+58\,q^{18}-74\,q^{16}+\nn\\&+103\,q^{14}-121\,q^{12}+155\,q^{10}-
175\,q^{8}+208\,q^{6}-224\,q^{4}+\nn\\&+247\,q^{2}+247\,q^{-2}-224
\,q^{-4}+208\,q^{-6}-175\,q^{-8}+155\,q^{-10}-\nn\\&-121\,q^{-12}+
103\,q^{-14}-74\,q^{-16}+58\,q^{-18}-36\,q^{-20}+25\,q^{-22}
-\nn\\&-12\,q^{-24}+7\,q^{-26}-2\,q^{-28}+q^{-30}-242\nn\\
A^{4}&
-q^{26}+q^{24}-3\,q^{22}+3\,q^{20}-5\,q^{18}+q^{16}-q^{
14}-5\,q^{12}+8\,q^{10}-18\,q^{8}+16\,q^{6}-\nn\\&-21\,q^{4}+20\,{q
}^{2}+20\,q^{-2}-21\,q^{-4}+16\,q^{-6}-18\,q^{-8}+8\,q^{-10}
-5\,q^{-12}-q^{-14}+q^{-16}-\nn\\&-5\,q^{-18}+3\,q^{-20}-3\,q^{-
22}+q^{-24}-q^{-26}-26\nn\\
A^{6}&
q^{20}-2\,q^{18}+5\,q^{16}-9\,q^{14}+15\,q^{12}-20\,q^{10}
+27\,q^{8}-32\,q^{6}+38\,q^{4}-40\,q^{2}-40\,q^{-2}+\nn\\&+38\,q^
{-4}-32\,q^{-6}+27\,q^{-8}-20\,q^{-10}+15\,q^{-12}-9\,q^{-14
}+5\,q^{-16}-2\,q^{-18}+q^{-20}+42\nn\ee
}
\section*{$\boxed{8_{10}}$}

\vspace{-1.3cm}
{\footnotesize
\be
A^{-6}&
-q^{20}+2\,q^{18}-6\,q^{16}+9\,q^{14}-16\,q^{12}+20\,q^{10
}-30\,q^{8}+32\,q^{6}-39\,q^{4}+39\,q^{2}+39\,q^{-2}-\nn\\&-39\,q
^{-4}+32\,q^{-6}-30\,q^{-8}+20\,q^{-10}-16\,q^{-12}+9\,q^{-
14}-6\,q^{-16}+2\,q^{-18}-q^{-20}-47\nn\\
A^{-4}&
q^{26}-q^{24}+4\,q^{22}-q^{20}+4\,q^{18}+6\,q^{16}-5\,q^
{14}+30\,q^{12}-34\,q^{10}+70\,q^{8}-\nn\\&-72\,q^{6}+119\,q^{4}-
106\,q^{2}-106\,q^{-2}+119\,q^{-4}-72\,q^{-6}+70\,q^{-8}-34
\,q^{-10}+\nn\\&+30\,q^{-12}-5\,q^{-14}+6\,q^{-16}+4\,q^{-18}-q^{
-20}+4\,q^{-22}-q^{-24}+q^{-26}+132\nn\\
A^{-2}&
-q^{30}+2\,q^{28}-8\,q^{26}+11\,q^{24}-26\,q^{22}+31\,q^{
20}-58\,q^{18}+58\,q^{16}-\nn\\&-95\,q^{14}+80\,q^{12}-124\,q^{10}+
87\,q^{8}-129\,q^{6}+76\,q^{4}-127\,q^{2}-\nn\\&-127\,q^{-2}+76\,{q
}^{-4}-129\,q^{-6}+87\,q^{-8}-124\,q^{-10}+80\,q^{-12}-95\,q
^{-14}+58\,q^{-16}-\nn\\&-58\,q^{-18}+31\,q^{-20}-26\,q^{-22}+11\,q
^{-24}-8\,q^{-26}+2\,q^{-28}-q^{-30}+68\nn\\
1&
2\,q^{30}-4\,q^{28}+14\,q^{26}-23\,q^{24}+51\,q^{22}-72\,q
^{20}+129\,q^{18}-\nn\\&-159\,q^{16}+242\,q^{14}-273\,q^{12}+371\,q
^{10}-377\,q^{8}+476\,q^{6}-\nn\\&-454\,q^{4}+530\,q^{2}+530\,q^{-2
}-454\,q^{-4}+476\,q^{-6}-377\,q^{-8}+\nn\\&+371\,q^{-10}-273\,q^{-
12}+242\,q^{-14}-159\,q^{-16}+129\,q^{-18}-72\,q^{-20}+\nn\\&+51\,q
^{-22}-23\,q^{-24}+14\,q^{-26}-4\,q^{-28}+2\,q^{-30}-474\nn\\
A^{2}&
-q^{30}+2\,q^{28}-8\,q^{26}+14\,q^{24}-33\,q^{22}+49\,q^{
20}-89\,q^{18}+119\,q^{16}-\nn\\&-186\,q^{14}+222\,q^{12}-303\,q^{
10}+333\,q^{8}-410\,q^{6}+414\,q^{4}-\nn\\&-471\,q^{2}-471\,q^{-2}+
414\,q^{-4}-410\,q^{-6}+333\,q^{-8}-303\,q^{-10}+\nn\\&+222\,q^{-12
}-186\,q^{-14}+119\,q^{-16}-89\,q^{-18}+49\,q^{-20}-33\,q^{-
22}+\nn\\&+14\,q^{-24}-8\,q^{-26}+2\,q^{-28}-q^{-30}+444\nn\\
A^{4}&
q^{26}-q^{24}+4\,q^{22}-5\,q^{20}+11\,q^{18}-13\,q^{16}+26
\,q^{14}-31\,q^{12}+50\,q^{10}-56\,q^{8}+\nn\\&+77\,q^{6}-78\,q^{
4}+95\,q^{2}+95\,q^{-2}-78\,q^{-4}+77\,q^{-6}-56\,q^{-8}+50
\,q^{-10}-31\,q^{-12}+\nn\\&+26\,q^{-14}-13\,q^{-16}+11\,q^{-18}-5
\,q^{-20}+4\,q^{-22}-q^{-24}+q^{-26}-88\nn\\
A^{6}&
-q^{20}+2\,q^{18}-5\,q^{16}+9\,q^{14}-15\,q^{12}+20\,q^{10
}-27\,q^{8}+32\,q^{6}-38\,q^{4}+40\,q^{2}+40\,q^{-2}-\nn\\&-38\,q
^{-4}+32\,q^{-6}-27\,q^{-8}+20\,q^{-10}-15\,q^{-12}+9\,q^{-
14}-5\,q^{-16}+2\,q^{-18}-q^{-20}-42\nn\ee
}

\end{document}